\documentclass[12pt]{iopart}%
\input epsfig
\usepackage{iopams}
\usepackage{amsfonts}
\usepackage{amssymb}
\usepackage{graphicx}%
\def\beqra{\begin{eqnarray}} \def\eeqra{\end{eqnarray}}
\def\beqast{\begin{eqnarray*}} \def\eeqast{\end{eqnarray*}}
\def\beq{\begin{equation}}      \def\eeq{\end{equation}}
\def\be{\begin{enumerate}}   \def\ee{\end{enumerate}}

\def\gam{\gamma}
\def\Gam{\Gamma}
\def\la{\lambda}

\def\La{\Lambda}
\def\si{\sigma}
\def\Si{\Sigma}

\def\om{\omega}

\def\pa{\partial}

\def\cg{{\cal G}}

\def\raisenot{\raise .5mm\hbox{/}}
\def\nota{\ \hbox{{$a$}\kern-.49em\hbox{/}}}
\def\notA{\hbox{{$A$}\kern-.54em\hbox{\raisenot}}}
\def\notb{\ \hbox{{$b$}\kern-.47em\hbox{/}}}
\def\notB{\ \hbox{{$B$}\kern-.60em\hbox{\raisenot}}}
\def\notc{\ \hbox{{$c$}\kern-.45em\hbox{/}}}
\def\notd{\ \hbox{{$d$}\kern-.53em\hbox{/}}}
\def\notbd{\ \hbox{{$D$}\kern-.61em\hbox{\raisenot}}} %big D
\def\note{\ \hbox{{$e$}\kern-.47em\hbox{/}}}
\def\notk{\ \hbox{{$k$}\kern-.51em\hbox{/}}}
\def\notp{\ \hbox{{$p$}\kern-.43em\hbox{/}}}
\def\notq{\ \hbox{{$q$}\kern-.47em\hbox{/}}}
\def\notW{\ \hbox{{$W$}\kern-.75em\hbox{\raisenot}}}
\def\notz{\ \hbox{{$Z$}\kern-.61em\hbox{\raisenot}}}
\def\notpa{\hbox{{$\partial$}\kern-.54em\hbox{\raisenot}}}

\def\fo{\hbox{{1}\kern-.25em\hbox{l}}}  %raised one

\def\tr{{\rm Tr}}
\def\trn{{\rm tr}_{_{\left(N\right)}}~}
\def\tr2n{{\rm tr}_{_{\left(2N\right)}}~}
\def\rmtr{{\rm tr}}
\def\dgg{\dagger}

\def\gmn{\cg_{\mu\nu}}

\def\smn{\Sigma_{\mu\nu}}
\def\pdgp{\phi^{\dagger}\phi}

\begin{document}
\title[Non-Hermitean RMT: Planar Diagrams and the ``Single-Ring'' Theorem ]
{Non-Hermitean Random Matrix Theory: Summation of Planar Diagrams, the 
``Single-Ring'' Theorem and the Disk-Annulus Phase Transition}
\author{Joshua Feinberg$^{a,b,c}$}

\begin{abstract}
I review aspects of work done in 
collaboration with A. Zee and R. Scalettar \cite{fz1,fz2,fsz} on complex 
non-hermitean random matrices. I open by explaining why the bag of 
tools used regularly in analyzing hermitean random matrices cannot be applied 
directly to analyze non-hermitean matrices, and then introduce the 
Method of Hermitization, which solves this problem. Then, for rotationally 
invariant ensembles, I derive a master equation for the average density of 
eigenvalues in the complex plane, in the limit of infinitely large matrices. 
This is achieved by resumming all the planar diagrams which appear in the 
perturbative expansion of the hermitized Green function. Remarkably, this 
resummation can be carried {\em explicitly} for any rotationally invariant 
ensemble.
I prove that in the limit of infinitely large matrices, the shape of 
the eigenvalue distribution is either a disk or an annulus. This is the 
celebrated ``Single-Ring'' Theorem. Which of these shapes is realized is 
determined by the parameters (coupling constants) which determine the ensemble.
By varying these parameters a phase transition may occur between the two 
possible shapes. I briefly discuss the universal features of this transition. 
As the analysis of this problem relies heavily on summation 
of planar Feynman diagrams, I take special effort at presenting 
a pedagogical exposition of the diagrammatic method, which some readers 
may find useful. 
\end{abstract}

\address{ $^a$Physics Department, University of Haifa at Oranim, Tivon 36006, 
Israel~\\ 
$^b$Physics Department, Technion, Israel Institute of Technology, Haifa 32000,
Israel~\\
 $^c$This is an expanded version of an invited talk at the 4th International 
Workshop on Pseudo-Hermitean Operators in Quantum Physics, Stellenbosch, 
South Africa, November 2005.~}
\ead{joshua@physics.technion.ac.il}

\section{Introduction}

Non-hermitean random matrices have interesting and important applications
in various physical problems. Examples for such applications are
the phase diagram of QCD\cite{qcd, zahed}, nuclear decay and resonances in 
multichannel scattering in chaotic and disordered systems \cite{nuclear,
nuclear1}, neural networks \cite{nuclear1},
and asymmetric random hopping matrices - the Hatano-Nelson model 
\cite{hn} - with applications to pinning of magnetic flux 
lines in high temperature superconductors and also as a new tool for studying 
Anderson localization in disordered conductors. Studying Anderson localization
by means of non-hermitean matrices is also one of the reasons for 
the very recent growing interest in studying the statistics of resonance 
widths and delay times in random media \cite{nuclear,resonances}. 
Our final example is diffusion in random velocity fields \cite{chalker}, which 
gives rise to random Fokker-Planck operators. 
For a nice recent overview of random matrix theory and its applications, 
including non-hermitean matrices, see \cite{encyclopedia}.

\subsection{The Difficulty Posed by Non-Hermitean Matrix Ensembles}
Let us recall first a few basic facts about hermitean random matrices. 
A basic tool in studying hermitean random matrices $\phi$
(henceforth all matrices will be taken to be $N\times N$ with $N$ tending to
infinity unless otherwise specified) is the Green's function defined by
\beq
G(z)=\langle {1\over N} \rmtr {1\over z-\phi}\rangle
\label{gz}
\eeq
where $\langle\cdots\rangle$ denotes averaging over the random distribution 
from which the matrices $\phi$ are drawn. Diagonalizing $\phi$ by a unitary 
transformation we have
\beqast
G(z)=\langle {1\over N} \sum_{k=1}^N {1\over z-\lambda_k}\rangle
\eeqast
where the $N$ real numbers $\{\lambda_k\}$ are the eigenvalues of $\phi$.
Thus, $G(z)$ is a meromorphic function with poles at the eigenvalues of 
$\phi$. In the large $N$ limit, the poles merge into a cut (or several cuts)
on the real axis\footnote{More precisely, this is really a statement
on the eigenvalues of an  {\em individual} realization of  $\phi$ in the 
large $N$ limit (assuming that all eigenvalues are typically 
confined in a finite segment, independent of $N$). The averaging process 
smudges the eigenvalues into cuts already at finite $N$.}. Powerful 
theorems from the theory of complex-analytic functions can then be brought 
to bear to the problem of determining $G(z)$ \cite{BIPZ}. All of this is 
well-known. 

In contrast, for a non-hermitean matrix, the eigenvalues invade the 
complex plane.  For example, for non-hermitean matrices $\phi$ generated 
with the probability $P(\phi)={1\over Z}e^{-N \rmtr \phi^{\dagger}\phi}$,  
Ginibre determined long ago \cite{ginibre} that in the large-$N$ limit, the 
eigenvalues are uniformly distributed over a disk of radius unity in the 
complex plane. The Green's function corresponding to this ensemble is 
\beqra
G(z, z^*)=\langle {1\over N} \rmtr {1\over z-\phi}\rangle = 
\left\{\begin{array}{cc} z^*~~~ & |z| <1 \\{} & {}\\
{1\over z}~~~ & |z|>1 \end{array}\right.\,,
\label{ginibre}
\eeqra
and it is obviously not analytic inside the unit disk. 
In general, the eigenvalues of a non-hermitean matrix taken from some
probability ensemble, will occupy, on average, a two-dimensional domain 
${\cal D}$ of the complex plane, where the Green's function is non-analytic. 
Thus, we lose the powerful aid of analytic function theory in analyzing these
ensembles.

In the theory of hermitean random matrices, a powerful
method consists of the Feynman diagrammatic expansion \cite{diagrams}
in which one expands $G(z)=\sum_{k} \langle \rmtr \phi^k \rangle/z^{k+1}$
as a series in $1/z$, the `` bare quark propagator." This method is
clearly no longer available in studying non-hermitean random matrices. The
knowledge of $G(z)$ as a series in $1/z$ can no longer tell us anything
about the behavior $G(z)$ for small $z$, as the simple example (\ref{ginibre}) 
above makes clear. The eigenvalues fill a two-dimensional region rather 
than a one-dimensional region, as is the case for hermitean matrices.

A renormalization group inspired method used in \cite{rg} implicitly 
involves an expansion in $1/z$ and is thus also not available for dealing 
with non-hermitean matrices without suitable further developments.

\subsection{Some Basic Formalism and Notations}

I shall now derive a couple of standard formulas which play a central role 
in studying the eigenvalue distribution of random non-hermitean matrices.
The derivation is straightforward (see, e.g., \cite{nuclear1} and Section 2 
in \cite{fz1}, and references therein), and is based on the 
identities\footnote{Our notations are standard: for $z=x+iy$ we define 
$\pa\equiv {\partial\over \partial z} = {1\over 2}\left({\partial\over 
\partial x} -i{\partial\over \partial y}\right)$ so that $\partial z=1$. 
Clearly, $\partial z^*=0$. Similarly, we define $\pa^{*}\equiv {\pa\over 
\pa z^*} = {1\over 2} \left({\pa\over\pa x} + i{\pa\over\pa y}\right)\,,$ 
so that $\pa^{*}~z^*=1$.}
$\pa~(1/z^*)=\pa^{*}~(1/z)=\pi\delta(x)\delta(y)$
and $\partial\pa^{*} ~\log(z z^*)=\pi\delta(x) \delta (y)\,.$

We are now ready to deal with the average density of eigenvalues of a 
non-hermitean matrix $\phi$. We diagonalize the matrix by a similarity
transformation\footnote{Non-diagonalizable matrices constitute a set of 
measure zero in the ensemble.} 
\beq\label{similarity}
\phi=S^{-1} \Lambda S
\eeq
where $\Lambda$ denotes a diagonal matrix with elements 
$\lambda_i$, $i=1,..., N$. Taking the hermitean
conjugate we have $\phi^{\dagger}=S^{\dagger} \Lambda^*
S^{-1 \dagger}$. By definition, the density 
of eigenvalues is 
\beq\label{rho1}
\rho(x,y)=\langle {1\over N} \sum_i \delta(x-\Re \lambda_i)~ \delta (y- \Im
\lambda_i)\rangle\,.
\eeq
From the identities mentioned above, we obtain that  
$\pa^*~\rmtr~(z-\La)^{-1} \equiv \pa^*~\rmtr~(z-\phi)^{-1} = 
\pi\sum_i \delta(x-\Re\la_i)~ \delta (y-\Im\la_i)$. Thus, using the identity 
$\det (z-\La)(z^*-\La^*) = \det (z-\phi)(z^*-\phi^\dgg)$
we may express $\rho$ in terms of $\phi$, and in a manner which is manifestly symmetric in $z$ and $z^*$, as 
\beq\label{rho}
\rho(x,y) = {1\over\pi} \partial\pa^{*}~\langle {1\over N} \rmtr~\log~(z-\phi)
(z^*-\phi^\dgg)\rangle\,.
\eeq
This expression involves the logarithm inside the average, as well as two 
derivatives. Thus, unless one has a simple way of calculating the average in 
(\ref{rho}) ({\em e.g.,} the replica method in the case of Gaussian ensembles 
\cite{nuclear1}), (\ref{rho}) is not a practical expression for $\rho(x,y)$. 
In many cases it is easier to calculate the Green's function
\beq\label{greens}
G(z,z^*)=\langle {1\over N} \rmtr {1\over z-\phi}\rangle = \langle {1\over N}
\sum_i {1\over (x-x_i) +
i(y-y_i)}\rangle\,,
\eeq
(with $\lambda_i= x_i+iy_i$) than to do the average 
$\langle {1\over N} \rmtr~\log~(z-\phi)(z^*-\phi^\dgg)\rangle\,.$ 
Then, from (\ref{rho1}) and from one of the identities mentioned above,
we have
\beq\label{rho11}
\rho(x,y) = {1\over\pi} \pa^{*}~G(z, z^*)\,,
\eeq
which is a simpler representation of $\rho(x,y)$ than (\ref{rho}).

These two representations for $\rho(x,y)$ and the relation between them can 
be interpreted by recognizing (\ref{rho}) as a two dimensional Poisson 
equation for the electrostatic potential $-{1\over 2} \langle {1\over N}~ 
\rmtr~\log~(z-\phi)(z^*-\phi^\dgg)\rangle$ created by the charge density 
$\rho(x,y)$. This connection between eigenvalue distributions of complex 
matrices and two-dimensional electrostatics has been long known in the 
literature \cite{nuclear1}. Continuing along this line, consider the Green's 
function $G(z,z^*)$ in (\ref{greens}).
If we define the electric field ${\bf E} = (\Re G, -\Im G)$ then
from the definition of $G(z,z^*)$ we have 
\beq\label{electric}
{\bf E} ({\bf x}) = \int d^2 y~ \rho({\bf y}) {{\bf x}-{\bf y}\over 
|{\bf x}-{\bf y}|^2}
\,
\eeq
and thus (\ref{rho11}) is simply the statement of Gauss' law for this 
electrostatic problem.

\subsubsection{A Quaternionic Interlude and Speculation}

I end this section with a speculative remark, made for the first time 
in \cite{fz1}. As was pointed out earlier, in the theory of random
hermitean matrices it is well known that it is much easier to work with
$G(z)=\langle {1\over N} \rmtr {1\over z-\phi}\rangle$ rather than with the 
density of eigenvalues $\rho(\mu)$ directly, since the power of analytic 
function can be brought to bear on $G(z)$. Of course, one can go from 
$\rho(x)$ to $G(z)$ with the identity 
\beq\label{plemelj}
{1\over\pi}~\Im{1\over x-i\epsilon}=
{1\over\pi}{\epsilon\over x^2+\epsilon^2}\rightarrow \delta(x)\,.
\eeq

Here, for random non-hermitean matrices our formula (\ref{rho}) gives the
density of eigenvalues $\rho(x,y)$ directly, but this formula is awkward
to work with because of the logarithm. Its direct analog in the theory of
random hermitean matrices would be $\rho(x)={\partial\over\partial x}
\langle {1\over N} \rmtr \theta(x-\phi)\rangle$, which would also be
extremely awkward to work with. It would thus be desirable to define a
quantity analogous to $G(z)$ and develop a method for calculating it.
Reasoning along these lines we are led to an attempt to write a function
of a quaternionic variable, in the same way that going from $\rho(x)$ to
$G(z)$ we went from a function of a real variable to a function of a
complex variable. Indeed, the obvious analog of (\ref{plemelj}) is easy to
find, namely, 
\beqra\label{quatplemelj} &&{1\over 2 j}~ \left[ {1\over
(z-j\epsilon )|z-j\epsilon |} + i {1\over (z-j\epsilon)|z-j\epsilon | }~i
\right]\nonumber\\{}\nonumber\\ &&={\epsilon\over |z-j\epsilon |^3}=
{\epsilon\over \left(x^2+y^2+\epsilon^2\right)^{3/2}}\rightarrow
2\pi\delta(x)~\delta(y)\,. 
\eeqra 
Here $z=x+iy$ is a complex number, $\epsilon$ is a small positive real number 
and $\{1,i,j,k\}$ is the standard basis of the quaternion 
algebra\footnote{ One may of course write an equivalent formula with $j$ 
replaced everywhere by the third quaternion basis element $k$. Also, 
$\epsilon$ may be taken complex.}. (The absolute value of a quaternion is 
defined by $|a+ib+jc+kd|\equiv \sqrt{a^2+b^2+c^2+d^2}$.)

Unfortunately, (\ref{quatplemelj}) as it stands, is less useful than 
(\ref{rho}), because it leads to the quaternionic Green's function $G(q)$
($q$ being a quaternionic variable)
\beq\label{quatG}
G(q) = \langle {1\over N} \sum_i {1\over (q-\lambda_i) | q-\lambda_i |}\rangle
\eeq
which involves the absolute value operation explicitly, and thus cannot be 
written as a simple trace like (\ref{gz}). Note, however, that if we split 
$q$ quite generally into $q=z+j w$ ($z,w$ being ordinary complex variables), 
then $G(z+j w)$ is manifestly non-analytic in $z$ as $w\rightarrow 0$, even 
before taking the average.

This was posed in \cite{fz1} as an interesting problem,
namely to find a useful quaternionic generalization of (\ref{gz}) for random 
non-hermitean matrices, and to develop the analogue of the work of 
Br\'ezin et al. \cite{BIPZ} associated with it. Unfortunately, this problem is
still open, at the moment of transcribing this talk. 

The rest of this paper is organized as follows: In Section 2 I present 
the Method of Hermitization \cite{fz1, zahed, efetov, chalker} by which 
the problem of determining the eigenvalue density of random non-hermitean 
matrices can be reduced to the more familiar problem of determining the 
eigenvalue density of random hermitean matrices. In Section 3 I explain how
this method was applied in \cite{fz2} to derive a master formula for the 
density of eigenvalues of non-hermitean random matrices taken from a large 
class of rotationally invariant non-gaussian probability ensembles. This 
section contains a detailed pedagogical exposition of the diagrammatic 
method and its large-$N$ behavior, which is the basis for the discussion 
in that section. In Section 4 I present numerical simulations of the quartic
ensemble, carried in \cite{fsz}, and demonstrate that they fit well with the
theoretical predictions of the formalism developed in Section 3. In Section
5 I formulate and prove the ``Single Ring'' Theorem of \cite{fz2}, according 
to which, in the large-$N$ limit, the shape of the eigenvalue distribution 
associated with any of the ensembles studied in Section 3 is either a disk or 
an annulus. Finally, in Sections 6 and 7, following \cite{fsz}, I present 
universal properties of the disk and annular phases and of the transition 
between them, as well as nice explicit expressions for the location 
of the boundaries of the eigenvalue distribution and its boundary values. 
Unfortunately, for lack of space, I cannot cover the interesting topic 
of adding random non-hermitean matrices. The interested reader could read
about it in \cite{fz2}.

\section{The Method of Hermitization: Reduction to Random Hermitean Matrices} 
I have emphasized that a straightforward diagrammatic method is not allowed 
for non-hermitean matrices. Somewhat remarkably, it is possible to arrive at a 
diagrammatic method indirectly. I will show that the problem of determining 
the eigenvalue density of random non-hermitean matrices can be reduced to the 
problem of determining the eigenvalue density of random hermitean matrices, 
for which the diagrammatic method may be applied.
This is achieved by the {\em Method of Hermitization}, as it was dubbed in 
\cite{fz1}. Variants of this idea were presented independently in 
\cite{zahed, efetov, chalker}. (Here we follow the conventions of \cite{fz1}.)

We start with the representation $\rho(x,y) = {1\over\pi} \partial\pa^{*}~
\langle {1\over N} \rmtr~\log~(z-\phi)(z^*-\phi^\dgg)\rangle\,$
(namely, Eq. (\ref{rho}).) By standard manipulations we observe that 
$\langle\trn\log (z-\phi)(z^*-\phi^\dgg)\rangle =\langle\tr2n\log H\rangle 
- i\pi N$, where $H$ is the hermitean $2N\times 2N$ matrix Hamiltonian 
\beqra\label{H}
H=\left(\begin{array}{cc} 0~~~ & \phi-z\\{} & {}\\
\phi^\dgg-z^* & 0\end{array}\right)\,.
\eeqra
Thus, if we can determine
\beq\label{object}
F(\eta; z, z^*) = {1\over 2N} \langle \tr2n\log (\eta - H)\rangle
\eeq
we can determine $\rho(x,y)$.

The matrix $H$ is what is called a chiral matrix \footnote{The problem
of determining the Green's function of chiral matrices such as $H$ has
been discussed by numerous authors \cite{chiral, chiral1, ambjorn}.
(See also the paper by Feinberg and Zee cited in \cite{rg}.)}. Due 
to its block structure, it anticommutes with the ``$\Gamma_5$'' 
matrix $\left(\begin{array}{cc} 1~~~ 
& 0\\0~~ & -1\end{array}\right)$, with the immediate consequence
that its spectrum consists of pairs of eigenvalues with equal magnitudes and 
opposite signs. Namely, if $\xi$ is an eigenvalue of $H$, so is $-\xi$.

Consider now the propagator matrix associated with $H$, namely, 
\beqra
\cg_{\mu\nu} (\eta; z, z^*) = \langle\left({1\over \eta-H}\right)_{\mu\nu}
\rangle
\label{propagator}
\eeqra
where $\eta$ is a complex variable and the indices $\mu$ and $\nu$ run over 
all possible $2N$ values. 
Here we followed a common practice and borrowed some terminology from 
quantum chromodynamics: we may consider $\phi, \phi^{\dgg}$ as ``gluons" 
(in zero space-time dimensions), which interact with a $2N$ dimensional 
multiplet of ``quarks"  $\psi^\mu$, with a complex mass matrix 
(the bare ``inverse propagator")
\beqra\label{inverseprop}
\cg_0^{-1} =\left(\begin{array}{cc} \eta~~~ & z\\{} & {}\\
z^* & \eta\end{array}\right)
\eeqra
(expressed in terms of its $N\times N$ blocks).  $\gmn$ is thus the propagator
of these quarks in the fluctuating gluon field. The crucial point is that 
since $H$ is hermitean, $\gmn $ can be determined by the usual methods of 
hermitean random matrix theory. In particular, as we already
mentioned, the diagrammatic evaluation of (\ref{propagator}) is essentially 
the expansion of $\gmn$ in powers of $1/\eta$, with interaction vertices $H$. 
This is a well defined procedure for large $\eta$, and it converges to a 
unique function which is analytic in the complex $\eta$ plane, except for the 
cut (or cuts) along the real axis which contain the eigenvalues of $H$. After 
summing this series (and thus determining $\gmn (\eta; z, z^*)$ in closed 
form), we are allowed to set $\eta\rightarrow 0$ in (\ref{propagator}). 
Speaking colloquially, we may say that the crucial maneuver here is that 
while we cannot expand in powers of $z$, we can arrange for $z$ to 
``hitch a ride" with $\eta$ and at the end of the ride, throw $\eta$ away.

We can now calculate $\rho(x,y)$ by two different methods, each with its 
advantages. The first is to observe that 
\beqast
-{1\over H} =\left(\begin{array}{cc} 0~~~ & {1\over z^* - \phi^\dgg}\\{} & {}\\
{1\over z- \phi} & 0\end{array}\right)\quad . 
\eeqast
Thus, the quantity ${1\over z- \phi}$ is simply the lower left block 
of $\gmn(\eta=0; z, z^*)$. In other words, once we have $\gmn (\eta; z, z^*)$
we can set $\eta$ to zero and use (\ref{greens}) and (\ref{rho11})
to write
\beq\label{hermitrho}
\rho(x,y) = {1\over N\pi} \pa^* \tr2n \left[\left(\begin{array}{cc} 0~~~ & 
{\bf 1}_N\\0~~ & 0~\end{array}\right)\cg (0;z, z^*)\right]\,.
\eeq
 
An alternative is to take the trace of $\gmn$:
\beq\label{GH}
\cg(\eta; z, z^*) = {1\over 2N} \langle \tr2n~{1\over \eta - H}\rangle
= {\eta\over N}\langle \trn~{1\over \eta^2 - (z^*-\phi^\dgg)(z-\phi)}\rangle\,,
\eeq
from which one can determine (\ref{object}) (in terms of a dispersion 
integral), and therefore determine $\rho(x,y)$, as explained in \cite{fz1} 
in detail. I shall not pursue this possibility here. However, a few important
comments relating to (\ref{GH}) are in order. Observe from (\ref{GH}) that 
\beq\label{observe}
\cg(0-i0; z, z^*) = {i\pi\over N} \langle \trn \delta\left(\sqrt{(z^*-\phi^
\dgg)(z-\phi)}\right)\rangle
\eeq
and thus counts the (average) number of zero-eigenvalues of the positive 
semi-definite hermitean matrix $\sqrt{(z^*-\phi^\dgg)(z-\phi)}$. 
Thus, if (on the average) $\phi$ has no eigenvalues equal to $z$, or in other 
words, if the density of eigenvalues of $\phi$ vanishes at $z$, then  
\beq\label{boundary}
\cg(0-i0; z, z^*)= 0\,,
\eeq
independently of the large $N$ limit. In particular, the boundaries of the 
eigenvalue distribution of $\phi$ in the large $N$ limit, are the curve 
(or curves) in the complex plane which separate regions where 
$\cg(0; z, z^*)= 0$ from regions where $\cg(0; z, z^*) \neq 0$. As a matter 
of fact, one can infer the location of these boundaries even without an 
explicit knowledge of $\cg(0; z, z^*)$ and $G(z, z^*)$, by investigating 
the ``gap equation" for $\cg(0; z, z^*)$, namely, the trace of both sides of 
(\ref{propagator}), and then setting $\eta=0$ \cite{fz1}. The ``gap equation" 
is a polynomial in $\cg$ with real coefficients (which depend on 
$z$ explicitly as well as through $G(z, z^*)$, and the parameters which 
appear in the probability distribution for $\phi$.) Due to the chiral 
nature of $H$, this polynomial contains a trivial factor of $\cg(0; z, z^*)$ 
which we immediately factor out. Setting $\cg =0$ in the remaining factor we 
obtain an equation for the boundary. This is discussed in detail in \cite{fz1}.

It is worth emphasizing that the formalism developed here has nothing to
do with large $N$ as such. It is also totally independent of the form of 
the probability distribution $P(\phi, \phi^\dgg)$. The formalism is of course 
indifferent to the method one may choose to use to determine 
$\gmn(\eta;z,z^{*})$. 

To summarize, we have obtained a formalism - the ``Method of Hermitization" -
for reducing the problem of dealing with random non-hermitean matrices to the 
well-studied problem of dealing with random hermitean matrices. 
Given the non-hermitean matrix $\phi$, we study the hermitean matrix 
$H$ instead. By whatever method one prefers, once one has determined the 
quark propagator $\gmn$ (or its trace, the Green's function 
$\cg (\eta;z,z^{*})$), one can in principle obtain $\rho(x,y)$ using 
(\ref{hermitrho}) (or using the other method mentioned following 
(\ref{GH})). Whether that can be done in practice is of course another story. 
In the next section I shall present a large family of probability 
distributions $P(\phi, \phi^\dgg)$ for which it is possible to 
compute $\gmn(\eta; z, z^*)$ in closed form (in the large-$N$ limit) by a 
diagrammatic method. Therefore, for those models, the algorithm discussed 
above can in fact be followed all the way to the end. 

\section{Rotationally Invariant Non-Gaussian Ensembles}

The literature on random non-hermitean prior to \cite{fz2} has focused 
exclusively on Gaussian randomness, a prime example being Ginibre's 
work \cite{ginibre} on the probability distribution 
$P(\phi) =(1/Z) {\rm exp}~(-N\rmtr\phi^\dgg\phi)$. In \cite{fz2} 
it was shown that using the method of hermitization one can 
determine the density of eigenvalues of probability distribution of the form 
\beq\label{prob}
P(\phi) = {1\over Z} e^{-N\rmtr V(\phi^\dgg\phi)}\,,
\eeq
where $V$ is an arbitrary polynomial of its argument. Indeed, by a simple 
trick, it was shown in \cite{fz2} that one can obtain the desired density of 
eigenvalues with a minimal amount of work, by judiciously exploiting the 
existing literature on random hermitean matrices. In this section I explain
how this can be done. 

In a certain sense, the work presented in \cite{fz2} may be thought of as 
the analog of the work of Br\'ezin et al. for random hermitean matrices 
\cite{BIPZ}; they showed how the density of eigenvalues of hermitean matrices 
$\phi$ taken from the probability distribution $P(\phi) = 
(1/Z){\rm exp} [-N\rmtr V(\phi)]$ with $V$ an arbitrary polynomial can be 
determined, and not just for the Gaussian case studied by Wigner and others, 
in which $V=(1/2)\rmtr\phi^2$. An important simplifying feature of the 
analysis in \cite{BIPZ} is that $P(\phi)$ depends only on the eigenvalues 
of $\phi$, and not on the unitary matrix that diagonalizes it. 
In contrast, the probability distribution (\ref{prob}) for non-hermitean 
matrices depends explicitly on the $GL(N)$ matrix $S$ used to diagonalize 
$\phi=S^{-1} \Lambda S$, and $S$ does not decouple. Remarkably however, for 
the Gaussian $P(\phi)$, Ginibre \cite{ginibre} managed to integrate over $S$ 
explicitly and derived an explicit expression for the probability distribution
of the eigenvalues of $\phi$. Unfortunately, it is not clear how to 
integrate over $S$ and derive the expression for the eigenvalue probability 
distribution for non-Gaussian distributions of the form (\ref{prob}). 
In \cite{fz2} this serious difficulty  was circumvented by using the method 
of hermitization.

Due to the symmetry of $P(\phi)$ under the transformation 
$ \phi\rightarrow e^{i\alpha}\phi$, the density of eigenvalues is obviously 
rotational invariant. It is useful to introduce to circularly invariant 
density as 
\beq\label{radrho}
\rho(x,y)\equiv\rho(r)/2\pi\,.
\eeq
Rotational invariance thus leads to a simpler form of the defining formula 
(\ref{greens}) for $G(z, z^*)$, reproduced here as 
$$G(z,z^*)=\langle {1\over N} \rmtr {1\over z-\phi}\rangle =
\int d^2 x'~{\rho(x', y')\over z-z'}\,, $$
which reads
\beq\label{circular}
\gam(r) \equiv zG(z,z^*)=\int\limits_0^r r'd r'~\rho(r')\,,
\eeq
whence
\beq\label{rhocirc}
\rho(r) = {1\over r} {d\gam\over dr}\,.
\eeq
Clearly, the quantity $\gam(r)$ is a positive monotonically increasing 
function, which satisfies the obvious ``sum-rules" 
\beq\label{sumrules}
\gam(0)=0 \quad\quad {\rm and}\quad\quad \gam(\infty)=1\,.
\eeq
In particular, observe that the first condition in (\ref{sumrules}) insures
that no $\delta(x)\delta(y)$ spike arises in $\rho(x,y)$ when calculating it 
from (\ref{rho11}) with $G(z, z^*)$ given by (\ref{circular}), as it should be.
We are now ready to perform the diagrammatic expansion for probability 
distributions of the form (\ref{prob}).
 
\subsection{The Diagrammatic Expansion and its Resummation}

Consider the matrix Green's function $\gmn$ in (\ref{propagator}) and write 
it as 
\beqra\label{propagator1}
\gmn = \langle\left({1\over {\bf\cg}_0^{-1} - V}\right)_{\mu\nu}
\rangle
\,,
\eeqra
where ${\bf\cg}_0^{-1}$ is the inverse propagator defined in 
(\ref{inverseprop}) and $V$ is the random chiral matrix (the ``gluon field'')
\beqra\label{random}
V=\left(\begin{array}{cc} 0~~~ & \phi\\{} & {}\\
\phi^\dgg & 0\end{array}\right)\,.
\eeqra
For large complex $\eta$ we may perfom the 
perturbative expansion of $\gmn$ in powers\footnote{Obviously, only terms 
with an even number of $V$'s survive in (\ref{perturbative}) for probability 
distributions of the form (\ref{prob}).} of ${\bf\cg}_0$
\beq\label{perturbative}
\gmn = {\bf\cg}_{0\mu\nu} + 
\langle\left({\bf\cg}_{0} V {\bf\cg}_{0}\right)_{\mu\nu}\rangle
+ \langle\left({\bf\cg}_{0} V {\bf\cg}_{0} V {\bf\cg}_{0}
\right)_{\mu\nu}\rangle + \cdots
\eeq
which converges to a unique matrix function, analytic in the $\eta$-plane,
cut along the appropriate segments on the real axis. This was our motivation 
to ``hermitize'' the problem to begin with. 

The diagrammatic method \cite{diagrams} provides a convenient and efficient 
way of representing the various terms which appear in the perturbation series 
(\ref{perturbative}). Before sketching the Feynman rules for this expansion,
some important symmetry considerations are in order. 

As was mentioned following (\ref{propagator}), $\gmn$ is essentially 
the propagator for the quarks $\psi^\mu \,(\mu=1,2,\ldots , 2N)$ in the 
fluctuating gluon field $V$. Let us split the $2N$ dimensional quark multiplet
$\psi^\mu$ into two $N$ dimensional ``flavors", $\psi = (u, d)$. We see then 
that the ``quark-quark-gluon" interaction is 
\beq\label{qqg}
\psi^\dgg V \psi = u^\dgg\phi d + d^\dgg\phi^\dgg u\,.
\eeq
This interaction term, and the probability distribution (\ref{prob})  
are manifestly invariant under the gauge transformation 
$\phi\rightarrow P\phi Q^\dgg, \phi^\dgg\rightarrow Q\phi^\dgg P^\dgg, 
u\rightarrow P u$ and $d\rightarrow Qd $, where $P$ and $Q$ are two 
independent unitary matrices, i.e., under the $U(N)\oplus U(N)$ subgroup of 
$U(2N)$. These transformations may be written more compactly as 
\beq\label{gauge}
V\rightarrow {\cal U} V {\cal U}^\dgg \,,\quad\quad  \psi\rightarrow 
{\cal U}\psi\,,\quad {\rm where}\quad {\cal U} =\left(\begin{array}{cc} 
P & 0\\0 & Q\end{array}\right)\,.
\eeq
We shall index quantities which transform under the first $U(N)$ in 
(\ref{gauge}) by Latin indices $i,j,\ldots $, and quantities 
which transform under the second $U(N)$ by indices from the beginning 
of the Greek alphabet $\alpha,\beta,\ldots $.(Indices from the middle 
of the Greek alphabet, $\mu,\nu,\ldots$ are reserved for $2N$ dimensional 
objects.) Thus, for example, we have $ u_i, d_\alpha $,and the mixed object 
$\phi_{i\alpha}$\,. 

\subsubsection{Feynman Rules and the Topological Nature of the ${1\over N}$
Expansion} 
Let us turn now to the Feynman rules for our diagrammatic expansion. 
Their presentation will be only sketchy and qualitative, since all that
will really matter to us in the discussion below is counting powers of $1/N$.
In fact, we are about to sum an infinite set of leading diagrams without ever 
computing any specific Feynman diagram. 

We start with the propagators. The bare quark propagators 
$\langle u_i u^*_j\rangle, \langle u_i d^*_\alpha\rangle, 
\langle d_\alpha u^*_i\rangle$ and $\langle d_\alpha d^*_\beta\rangle,$
are simply the four $N\times N$ blocks of ${1\over N} {\bf\cg}_{0}$. 
They are all of order ${1\over N}$. (Note that the object 
traced over in (\ref{hermitrho}) and (\ref{GH}) is essentially 
${1\over N}\gmn$. This trivial observation gives at least a 
practical justification for the factor ${1\over N}$ in the bare quark 
propagator. A deeper reason for this normalization of the quark 
propagator will be provided below.) These propagators are indicated in 
Fig.1 (b)-(e). The diagonal blocks of ${1\over N} {\bf\cg}_{0}$ correspond to 
quark propagation without flavor changing. The $uu$ propagator is indicated 
in  Fig.1(b) by a full line, and the $dd$ propagator is indicated in  Fig.1(c)
by a dashed line. The off-diagonal blocks correspond to quark propagation 
with flavor changing. 
The lines are all directed since the quarks are ``charged'' complex fields. 
Since all the blocks of ${1\over N}{\bf\cg}_{0}$ are proportional to the 
$N\times N$ unit matrix, color is preserved under quark propagation, whether 
flavor is changed or not. 
%%%%%%%%%%%%%%%%%%%%%%
\vspace{24pt}
\par
\hspace{0.5in} \epsfbox{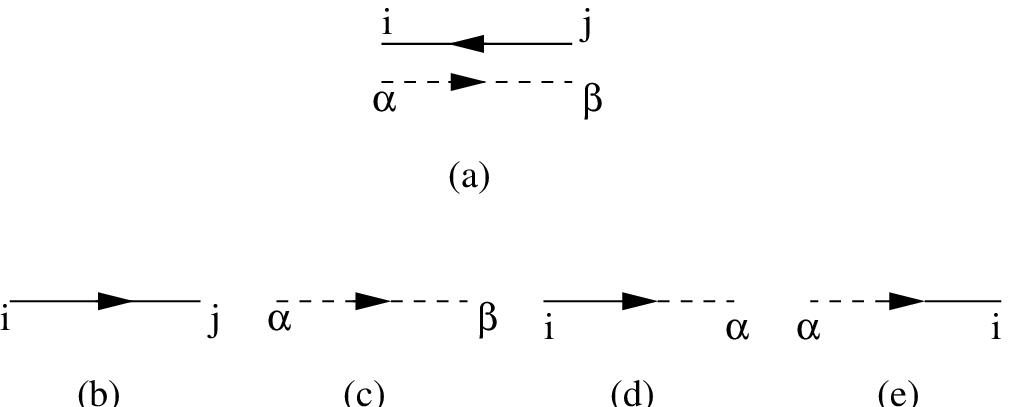}
\par
\baselineskip 12pt
{\footnotesize {\bf Fig.~1:} The bare gluon propagator (a), and the four 
blocks of the bare quark propagator (b)-(e). All propagators are of order
${1\over N}$.}
\vspace{24pt}
\par
\baselineskip 20pt   
%%%%%%%%%%%%%%%%%%%%%
Consider now the ``pure glue'' action
\beq\label{potential}
N\rmtr V(\phi^\dgg\phi) = Nm^2 \rmtr \phi^\dgg\phi + Ng_4 \rmtr
(\phi^\dgg\phi)^2 + Ng_6 \rmtr (\phi^\dgg\phi)^3 + \cdots
\eeq
in the exponent of (\ref{prob}). By definition, it is proportional to $N$, 
and the ``gluon self-couplings'' $m^2, g_4, g_6, \ldots $ are all of order 
$N^0$. Thus, a typical matrix element in $\phi$, drawn from the ensemble 
defined by (\ref{prob}), is of order $1/\sqrt{N}$ (and arbitrary phase), 
which leads, by standard arguments, to the conclusion that typical 
eigenvalues of $\phi$ are of order $N^0$, as was already mentioned in the 
context of Ginibre's ensemble. 

The bare gluon propagator can be read-off the quadratic term in 
(\ref{potential}) in the usual manner. We find 
\beq\label{gluonprop}
\langle \phi_{i\alpha} \phi^*_{j\beta}\rangle_0 = {1\over N m^2} 
\delta_{ij}\delta_{\alpha\beta}\,.
\eeq
It is of order ${1\over N}$, as a direct consequence of the explicit $N$ 
dependence in (\ref{prob}). In order that the quark propagator have the 
same $N$-dependence as the gluon's, we scaled the quark propagator to be of 
order ${1\over N}$ as well. This behavior, together with the fact that 
the ``gluon self-couplings'' $m^2, g_4, g_6, \ldots $ are all of order 
$N^0$, will ensure that the model has a smooth behavior as 
$N\rightarrow\infty$, as will be seen in more detail below.

The bare gluon propagator is depicted in Fig.1(a). Following 
't Hooft \cite{thooft}, it is represented by a double line, 
simply because $\phi$ carries two indices. The flows along these lines
preserve both color and flavor, as is evident from (\ref{gluonprop}). 
The reason why the lines in Fig.1(a) run in opposite directions is easy
to understand. Observe that $\phi$ transforms under gauge transformations
as $u\otimes d^\dgg$. Thus, the bare propagation of $\phi$ is similar to 
a process in one end of which a $u$-quark is destroyed and a $d$-quark 
is created, and in the other end of which the opposite occurs. 

Let us discuss now the vertices. From the ``quark-quark-gluon'' interaction 
(\ref{qqg}) we see that a gluon $\phi$ always absorbs a $d$-quark and emits a 
$u$-quark, and $\phi^\dgg$ does the opposite. These two processes are 
described by the two quark-gluon vertices depicted in Fig.2(a),(b). Both 
vertices
are taken to be proportional to $N$, so that they be of the same order of 
magnitude as the gluon self-interaction vertices, such as the quartic
vertex with coupling $Ng_4$ shown in Fig.2(c) and the sextic vertex with 
coupling $Ng_6$ in Fig.2(d). Another way of seeing that the quark-gluon 
vertices should be of order $N$ is to note from (\ref{propagator1}) that 
with the ${1\over N}$ normalization of the quark propagator we have 
${1\over N}\gmn = \langle\left({1\over N{\bf\cg}_0^{-1} - NV}\right)_{\mu\nu}
\rangle $. 
%%%%%%%%%%%%%%%%%%%%%%
\vspace{24pt}
\par
\hspace{0.5in} \epsfbox{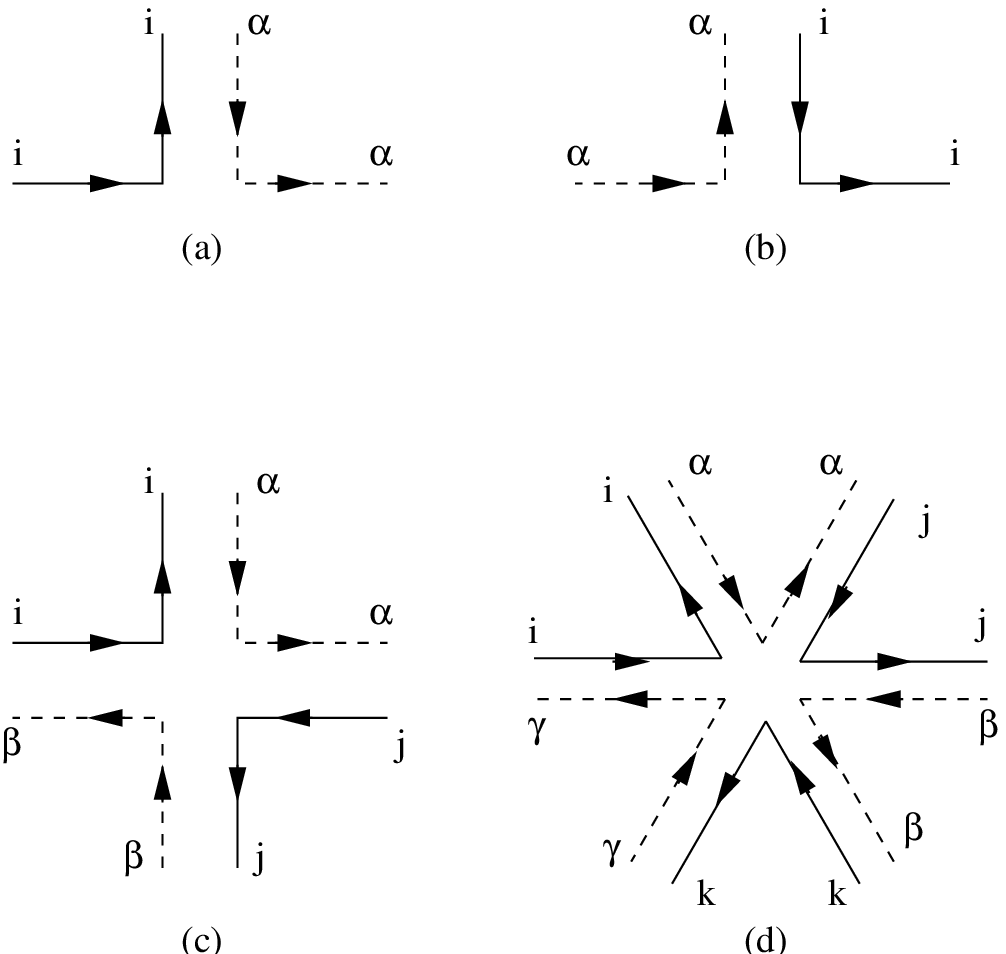}
\par
\baselineskip 12pt
{\footnotesize {\bf Fig.~2:} The two bare quark-gluon vertices (a),(b) and 
quartic- (c) and sextic- (d) gluon self-interaction vertices. All vertices
are of order $N$. }
\vspace{24pt}
\par
\baselineskip 20pt   
%%%%%%%%%%%%%%%%%%%%%
That all vertices scale in the same way with 
$N$, as do the propagators, is needed to obtain a smooth large-$N$ behavior. 
Note from Figs.2(a)-(d) that both color and flavor flow unobstructed 
through the various vertices. This is, of course, because all interaction 
terms are gauge invariant.

To summarize, all bare vertices are of order $N$, and all propagators 
are of order ${1\over N}$. 

Now, use the vertices and propagators defined above as building blocks of 
graphs - Feynman diagrams. The diagrams thus constructed are called 
double-line diagrams. 

In order to compute a certain observable, all
double-line diagrams consistent with that observable should be drawn, 
computed and added with the appropriate combinatoric factors. Note that when 
a closed index loop occurs in a diagram, its index should be summed over. 
Therefore, we should assign a factor $N$ to each index loop. 

We would like now to study the ${1\over N}$ behavior of these double-line 
diagrams, and show that it is determined by their topological 
properties \cite{thooft}. (An excellent discussion of this issue is given in 
\cite{coleman}, which I shall now follow with the appropriate modifications.)
Let us start, for simplicity, with connected ``vacuum to vacuum'' 
diagrams, i.e., connected double-line diagrams without external legs. 
(Such diagrams arise, for example, in the perturbative computation of the 
normalization factor $Z$ in (\ref{prob}). Their sum is essentially $\log Z$. 
The expansion (\ref{perturbative}) of the quark propagator is of course not 
the sum of vacuum diagrams.) Since 
there are no external legs, each index line in the diagram has to be a closed 
{\em oriented} loop. 
Obviously, there are two types of such loops - loops made of a full line 
and loops made of a dashed line. Let us attach a little surface - a polygon - 
to each of these index loops. Since there are two kinds of loops - there
are two kinds of such polygons. Thus, we can think of the double-line 
diagram in question as a tiled, triangulated (or more precisely - polygonated)
big two-dimensional surface, where any pair of edges of two different 
polygons, which share a gluonic 
double-line propagator, are identified. Since these two edges always run in
opposite directions (recall Fig.1(a)), the orientations of all polygons are 
the same, and the surface thus constructed is {\em orientable}
\footnote{This is all due to the fact that all lines here are directed. For 
real {\em asymmetric} matrices, we would obtain unoriented lines and 
therefore unoriented surfaces \cite{BN}, e.g., a Klein bottle.}.
Moreover, since the two lines which make a gluon propagator are of different 
types, and since index lines flow unobstructed through the vertices, any two 
neighboring polygons which share a gluonic edge must be of different types. 
In other words, our surface has a checkerboard structure\footnote{
Such checkerboard surfaces appear in certain models of two-dimensional 
quantum gravity. See e.g. \cite{ambjorn}.}. 
This surface may have edges, formed by quark lines. Also, it will generally 
be multiply connected - i.e., contain ``handles'' and internal holes.
An example of such a (simply connected) surface appears in Fig.3. 
%%%%%%%%%%%%%%%%%%%%%%
\vspace{24pt}
\par
\hspace{0.5in} \epsfbox{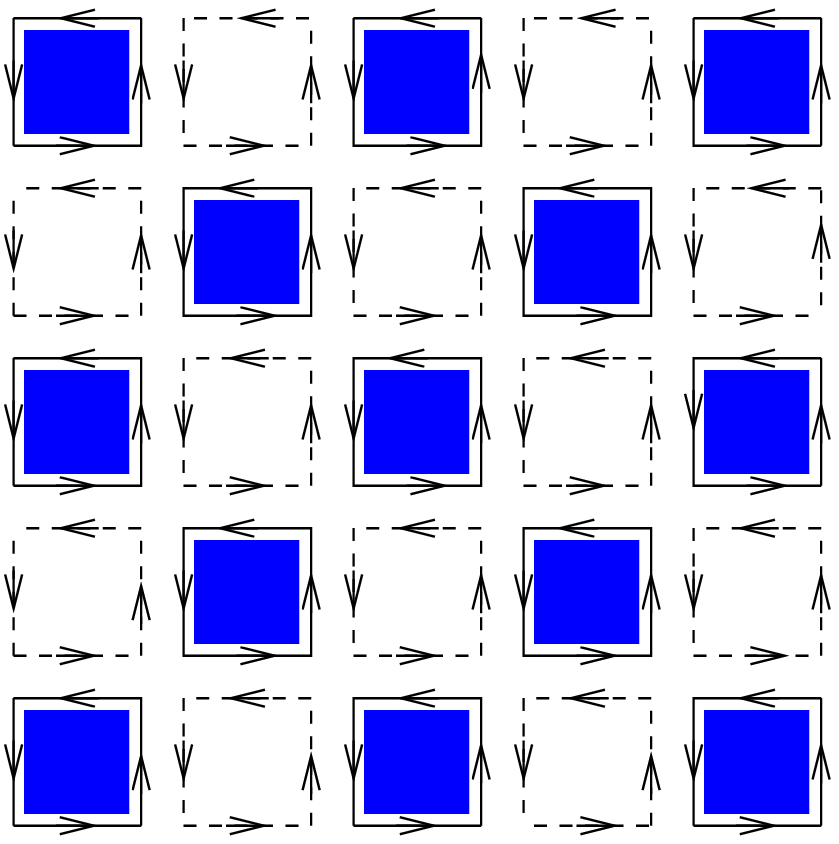}
\par
\baselineskip 12pt
{\footnotesize {\bf Fig.~3:} An example of a simply connected 
checkerboard surface which contains, in addition to quark-gluon 
vertices, only quartic gluon self-interaction vertices. There is a single 
quark loop which forms the edge. This diagram has 16 quark-gluon vertices 
plus 16 gluon quartic vertices ($V=32$), 25 index loops (faces, $F=25$), 
and 40 gluon propagators and 16 quark propagators (edges, $E=56$). 
As explained below (see (\ref{euler})), this diagram scales 
like $N$. }
\vspace{24pt}
\par
\baselineskip 20pt   
%%%%%%%%%%%%%%%%%%%%%
Counting the power of $N$ associated with such surface is easy. 
Assume the surface in question has $V$ vertices, $E$ edges and $F$ faces. 
Each vertex is an interaction vertex in the Feynman diagram, and thus 
contributes a factor $N$. Each edge is a propagator which carries a factor 
${1\over N}$, and each face is an index-loop carrying a factor $N$. 
All-in-all, the graph is proportional to $N^{F-E+V}\equiv N^\chi$, 
where
\beq\label{euler}
\chi = F-E+V 
\eeq
is the Euler character of the graph - a topological invariant of the graph.

As is well-known, every connected oriented two-dimensional surface is 
topologically equivalent to a sphere with an appropriate number of punctures 
(holes) and handles (or ``wormholes'') attached to it. For example, a torus is 
equivalent to a sphere with one handle attached, a disk - to a sphere with 
one puncture, a cylinder  - to a doubly punctured sphere, etc. 
Let us denote the number of handles (the so-called {\em genus} of the graph) 
by $G$, and the number of punctures 
(holes or boundaries) by $b$. Then a well-known theorem states that 
\beq\label{genus}
\chi = 2 - 2G - b\,.
\eeq
Thus, the leading vacuum diagrams in the limit $N\rightarrow\infty$ 
must have $G=b=0$, i.e., the topology of a sphere. They are thus 
proportional to $N^2$. Obviously, they cannot contain quark loops, which 
are unpaired index loops and thus correspond to holes in the surface 
($b\neq 0$). The last observation is more or less straightforward - 
one needs just one vertex factor to create either a pair of quarks or a 
triplet of gluons, but there are essentially $N$ times more gluon types to 
sum over than quarks.

By removing a randomly chosen face off our leading order graph, we can 
continuously stretch it open and project it onto the plane. For this reason, 
these leading vacuum graphs with spherical topology are also known as 
{\em planar} graphs. (We can also patch this planar graph back into a 
sphere, by including the point at infinity.)

Vacuum diagrams which depend of properties of quarks (i.e., on $\eta, z, z^*$ 
in our case) must contain at least one quark loop. It follows from 
(\ref{genus}) that the leading such graphs in the large-$N$ limit must 
contain precisely one quark loop (i.e. - a single hole) and no handles. 
Such graphs have the topology of a disk, and could thus be drawn on the
plane. They are planar graphs, like the pure glue graphs, with the sole 
difference between the two cases being that the outer boundary of that graph 
is that of the hole - i.e., the quark loop. There can be no quark loops in the 
{\em interior} of the graph in the large-$N$ limit.

The reader may identify the Feynman rules described above as those which 
correspond to the zero-dimensional ``quantum field theory'' with lagrangian 
\beq\label{lagrangian}
{\cal L} = N\rmtr V(\phi^\dgg\phi) + \psi^\dgg (N{\cal G}_0^{-1} - NV)\psi\,.
\eeq  
Diagrams which contribute in higher orders of the ${1\over N}$ expansion of 
this theory do contain quark loops in their interiors, i.e., correspond to 
surfaces with internal holes. Diagrams with such internal quark loops 
obviously {\em do not} appear in the expansion of 
${1\over N}\langle\gmn\rangle$ since the random matrix probability 
distribution (\ref{prob}) really does not contain any dynamical quarks. 
One method of suppressing these internal quark loops to any order in the 
${1\over N}$ expansion of (\ref{lagrangian}) is to apply the so-called 
``replica trick''\setcounter{footnote}{0}\footnote{That is to 
say, to replicate the quarks in (\ref{lagrangian}) into $n$ copies, with the 
same inverse bare propagator and coupling to the gluons. One then computes 
various observables in the replicated theory as a function of $n$, considered
as a complex variable, and then take the limit $n\rightarrow 0$. Each internal
quark loop costs a factor $n$ and is therefore suppressed in that limit.} 
\cite{replica}. Since here we are interested only in the 
$N\rightarrow\infty$ limit, namely, the leading order in the 
${1\over N}$ expansion, there is no need to invoke any replica considerations.

\subsubsection{Computation of ${1\over N}\gmn$ to Leading Order in 
${1\over N}$}

Obviously, by cutting open the quark loop at the boundary of our vacuum 
diagram at one point, and attaching color indices $\mu$ and $\nu$ at the two 
ends, we obtain a diagram which contributes to ${1\over N}\gmn$ in the 
expansion (\ref{perturbative}). Thus, to leading order in ${1\over N}$, all 
diagrams which contribute to ${1\over N}\gmn$ (and are therefore of order 
${1\over N}$) are planar diagrams, with the quark line running at the 
boundary, and all gluon double-lines are attached to it on the 
{\em same side}, without crossing each other. A couple of such diagrams
appear in Fig.4.
%%%%%%%%%%%%%%%%%%%%%%
\vspace{24pt}
\par
\hspace{0.5in} \epsfbox{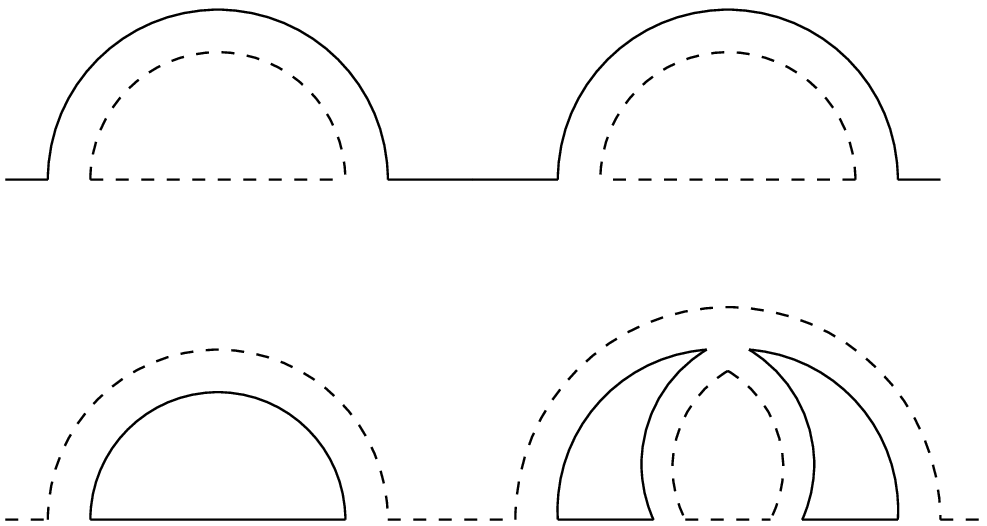}
\par
\baselineskip 12pt
{\footnotesize {\bf Fig.~4:} Two simple planar diagrams which contribute to 
${1\over N}\gmn$. The upper one contributes to the $uu$ block and the lower one
to the $dd$ block. Both diagrams are of order ${1\over N}$. Color indices and
arrows are suppressed. These diagrams are one-quark-reducible: their topology 
is such that they can be split into two disjoint pieces by cutting a single 
quark line.}
\vspace{24pt}
\par
\baselineskip 20pt   
%%%%%%%%%%%%%%%%%%%%%
I shall now explain how to sum all these planar diagrams systematically and 
thus obtain ${1\over N}\gmn$ in the limit $N\rightarrow\infty$.

The quark propagator (\ref{propagator}) is given by the Dyson-Schwinger 
equation in terms of the one-quark irreducible self-energy $N\smn$:
\beq\label{selfenergy}
{1\over N}\gmn =\left( {1\over N{\bf\cg}_0^{-1} - N{\bf\Sigma}}
\right)_{\mu\nu}\,.
\eeq
We emphasize that this equation always holds, and practically amounts to the 
definition of the one-quark irreducible self energy. Evidently, all planar 
self-energy diagrams are of order $N$\footnote{Their large-$N$ behavior is 
like that of the planar diagrams which contribute to  ${1\over N}\gmn$ with 
their two external legs amputated. Amputating two external legs amounts to
multiplication of the latter by $N^2$. Another way of looking at this is
to say that the self-energy should scale with $N$ as does the 
quark-quark-gluon vertex $NV$ which is combined with the inverse propagator 
in the averaged quantity in (\ref{propagator}).}. Thus, to leading order, 
$\smn$ is of order $N^0$. A couple of planar 
one-quark irreducible diagrams which contribute to $N\smn$ appear in Fig.5.
%%%%%%%%%%%%%%%%%%%%%%
\vspace{24pt}
\par
\hspace{0.5in} \epsfbox{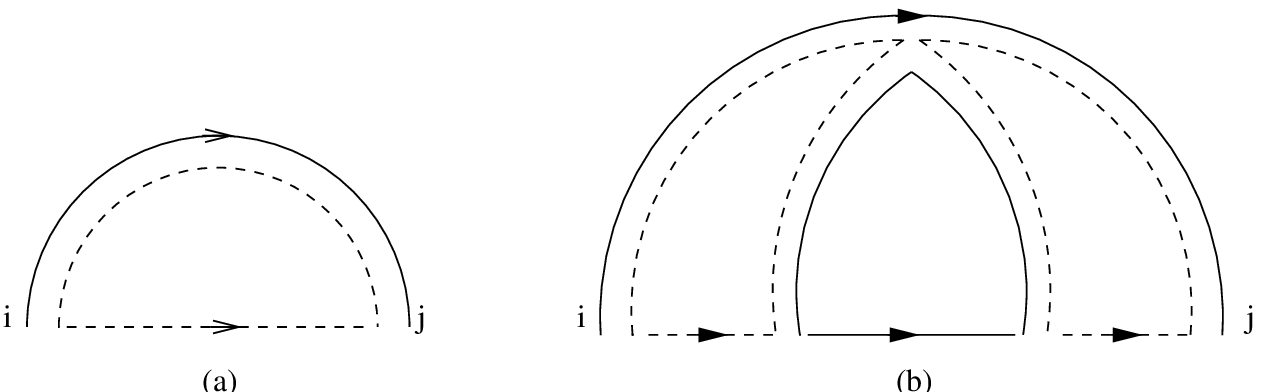}
\par
\baselineskip 12pt
{\footnotesize {\bf Fig.~5:} Two simple planar one-quark irreducible diagrams
which contribute to the $uu$ block of $N\smn$: a diagram with no gluon 
self-interactions (a), and a diagram with a quartic gluon self-interaction 
(b). Evidently, both diagrams are proportional to $\delta_{ij}$. Also, 
both diagrams are of order $N$.}
\vspace{24pt}
\par
\baselineskip 20pt   
%%%%%%%%%%%%%%%%%%%%%
For pedagogical clarity, we also display in Fig.6 a couple of non-planar 
self-energy diagrams which contribute in the next-to-leading order. 
%%%%%%%%%%%%%%%%%%%%%%
\vspace{24pt}
\par
\hspace{0.5in} \epsfbox{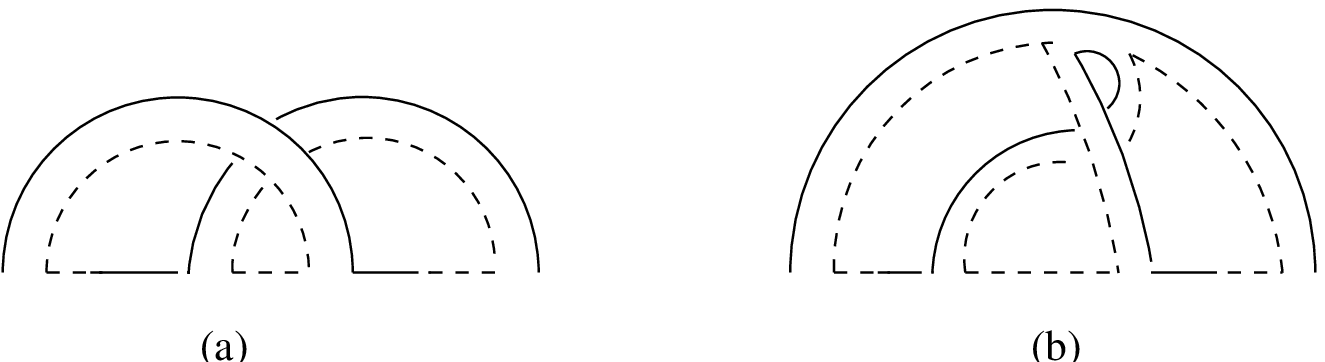}
\par
\baselineskip 12pt
{\footnotesize {\bf Fig.~6:} Two simple non-planar self-energy diagrams with 
a single crossing of gluon lines. Color indices and arrows 
are suppressed. The effect of this crossing is dramatic: 
in diagram (a) there are no index loops at all (as opposed to its (reducible) 
counterpart in the upper part of Fig.4 with two such loops), and in (b) there is only one 
index loop (as opposed to its planar counterpart in Fig.5(b) with three
such loops). These diagrams are both of order ${1\over N}$, i.e., down by a 
factor $N^{-2}$ with respect to the leading order.}
\vspace{24pt}
\par
\baselineskip 20pt   
%%%%%%%%%%%%%%%%%%%%%
The planar self-energy diagrams shown on Fig.5 are clearly proportional to the 
unit matrix. By carefully studying a few additional planar self-energy 
diagrams the reader can convince himself or herself that this is a general 
property: in these planar self-energy diagrams, there is always a unique 
peripheral index line, which emanates from the quark-quark-gluon vertex at one 
end of the diagram, and participates in several gluon propagators and flows 
unobstructed through a number of gluon self-interaction vertices, without 
crossing any other index line, all the 
way to the quark-quark-gluon vertex at the other end of the 
diagram. (A slightly different way of looking at this is to connect the
two ends of the diagram by a quark propagator and close it into a planar 
vacuum diagram. The point of gluing lies on a quark line, which is of course 
the {\em outer} edge of one of the polygons making the surface. The other 
edges of that polygon lie within the surface and combine to a directed
index line, which becomes the peripheral index line mentioned above upon
cutting the diagram open.) 
Thus, evidently, the off-diagonal blocks of $N\smn$ are null and its $uu$ and 
$dd$ blocks must be proportional to the unit matrix. Furthermore, the 
$u$ and $d$ quarks are completely equivalent. (We can switch the roles of 
$\phi$ and $\phi^\dgg$ in (\ref{prob}) and nothing would change.) Thus, the
$uu$ and $dd$ blocks of $\smn$ must be equal. In other words, for probability 
distributions of the form (\ref{prob})
\beq\label{Sigmamn}
\Sigma_{\mu\nu} = \Si(\eta; r) \delta_{\mu\nu}
\eeq
(where we recall that $r=|z|$.) Note that $\Si$ depends on $r$ and not 
separately on $z$ and $z^*$ because it may be expanded as an infinite sum of 
traces of $\cg_0$, each term of which is a function only of $\eta$ and $r$. 
Thus, from (\ref{selfenergy}),   
\beq\label{gmn}
\gmn = {1\over r^2 - (\eta-\Si)^2} ~
\left(\begin{array}{cc} \Si-\eta ~~~ & z\\{} & {}\\
z^* & \Si-\eta\end{array}\right)\,,
\eeq
leading to 
\beq\label{selfenergy1}
\cg \equiv {1\over 2N}\sum_\mu \cg_{\mu\mu} = {\Si - \eta  \over  r^2 - 
(\eta-\Si)^2}\,.
\eeq
The important point \cite{fz2, bluez} is that in the large-$N$ limit, the 
one-quark irreducible self-energy $N\Sigma$ can be written in terms of the 
cumulants $\Gamma_{2k}$ of $P(\phi) = (1/Z) e^{-N\rmtr V(\phi^\dgg\phi)}$ 
(Eq. (\ref{prob})), namely, the connected correlators involving 
$k$ $\phi$'s and $\phi^\dgg$'s. Thus, to leading order in ${1\over N}$, this
cumulant is the sum of all connected planar pure glue diagrams with $k$ 
external $\phi$'s and $\phi^\dgg$'s (with a prescribed index structure). 
By definition, $\Gamma_{2k}$ is a quantity of order $N$. (It could be thought 
of essentially as the ``dressed'' gluon self-interaction vertices $Ng_{2k}$.) 
This makes sense, since any of the diagrams contributing to $\Gamma_{2k}$ can 
be thought of as a planar, connected pure glue vacuum diagram 
(and thus of order $N^2$), cut open at $k$ places along its {\em peripheral} 
gluon propagators\footnote{These cuts must be made along peripheral gluon 
propagators since the gluon diagram in question must be connected, without 
crossings, to other parts of a bigger planar diagram.} in such a way that it 
remains connected (i.e., there must be at least one gluon self-interaction
vertex between consecutive cutting points), and then amputating the gluon 
legs thus formed. A moment's reflections reveals that this operation removes 
$k+1$ index loops (i.e., $k$ internal index loops as well as the outer 
perimeter loop) and also $k$ gluon propagators. Thus, it costs a factor of 
${1\over N}$, reducing the $N$-power counting of the diagram from 2 to 1.

Diagrammatically speaking, the $\Gamma_{2k}$ may be thought of as a proper
vertex of $k$ $\phi$'s and $\phi^\dgg$'s, i.e., upon embedding it in a
planar diagram it appears as a ``blob" out of which emanate $k$ $\phi$'s and 
$\phi^\dgg$'s in an alternate manner and without crossings, and which cannot 
be separated into two smaller blobs, with $k_1$ $\phi$'s and $\phi^\dgg$'s in 
one blob and $k_2$ $\phi$'s and $\phi^\dgg$'s in the other 
(with $k_1+k_2=k$ of course.)

The way $N\Si$ is determined by the cumulants $\Gamma_{2k}$ can be deduced by 
working out a few simple cases. For example, the contribution of $\Gamma_2$ 
is shown in Fig.7. 
%%%%%%%%%%%%%%%%%%%%%%
\vspace{24pt}
\par
%\hspace{0.5in} \epsfbox{talk-feynman-rules.eps}
\hspace{0.5in} \epsfbox{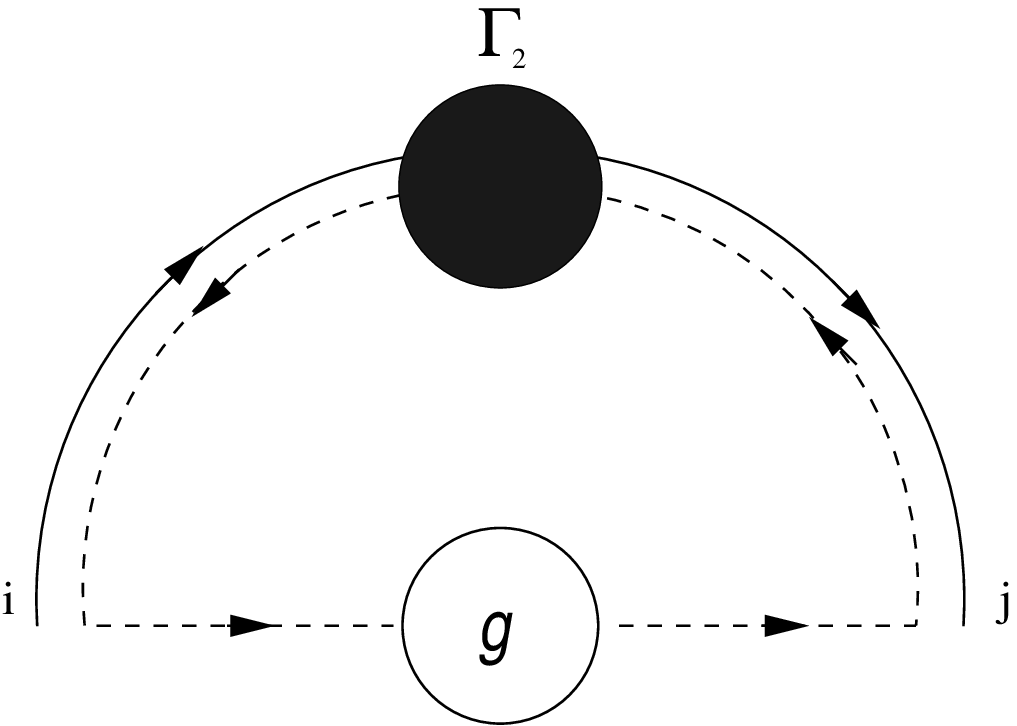}
\par
\baselineskip 12pt
{\footnotesize {\bf Fig.~7:} Contribution of the quadratic cumulant $\Gamma_2$
to self-energy.}
\vspace{24pt}
\par
\baselineskip 20pt   
%%%%%%%%%%%%%%%%%%%%%
Counting powers of $N$ in Fig.7 is as follows: $\Gamma_2$ is of order 
$N$. The ${1\over N}$ factors associated with the two gluon propagators 
poking out of the dark gluon blob representing $\Gamma_2$ are canceled 
against the factors of $N$ in the quark-gluon vertices. The horizontal line 
at the bottom is the {\em dressed} planar quark propagator, i.e., the sum of 
all planar diagrams contributing to ${1\over N}\gmn$, and is of order 
${1\over N}$. It is connected to the internal index line which 
emanates from the quark-quark-gluon vertex on the left and flows through
the gluon dark blob. On the way, that line participates in several gluon 
propagators and flows unobstructed through a number of gluon self-interaction 
vertices in any of the planar pure glue diagrams which make the blob, without 
crossing any other index line, all the way to the quark-quark-gluon vertex 
on the right. Thus, it preserves its color index, which must be summed over. 
Thus, the dressed quark propagator is traced over and produces a factor 
$\cg(\eta;r)$ of order $N^0$, as defined in (\ref{selfenergy1}). The overall 
contribution of Fig.7 to self-energy is therefore $\Gamma_2 \cg$ which is of 
order $N$. Evidently, the diagram shown in Fig.5(a) is the leading 
contribution to this quantity. It will be instructive to analyze the 
contribution of $\Gam_4$, the next cumulant, as well. Its contribution 
to self-energy is shown in Fig.8. 
%%%%%%%%%%%%%%%%%%%%%%
\vspace{24pt}
\par
%\hspace{0.5in} \epsfbox{talk-feynman-rules.eps}
\hspace{0.5in} \epsfbox{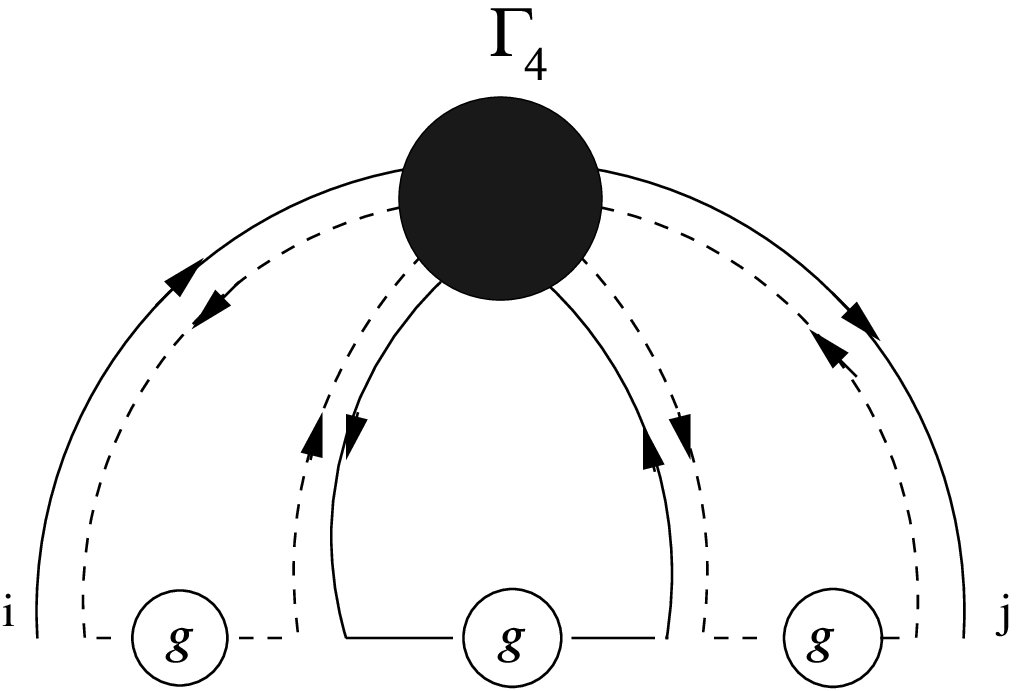}
\par
\baselineskip 12pt
{\footnotesize {\bf Fig.~8:} Contribution of the quartic cumulant $\Gamma_4$
to self-energy.}
\vspace{24pt}
\par
\baselineskip 20pt   
%%%%%%%%%%%%%%%%%%%%%
Counting powers of $N$ in Fig.8 is as follows: $\Gamma_4$ is of order 
$N$. The ${1\over N}$ factors associated with the four gluon propagators 
poking out of the dark gluon blob are canceled against the factors of $N$ in 
the quark-gluon vertices. Each one of the three dressed quark propagators at 
the bottom is connected to an internal index loop which flows through
the gluon dark blob. Thus it is traced over and produces a factor 
$\cg(\eta;r)$. The overall contribution of Fig.8 to self-energy is 
therefore $\Gamma_4 \cg^3$ which is of order $N$. Evidently, the diagram 
shown in Fig.5(b) is the leading contribution to this quantity.

The general pattern should be obvious\footnote{It should also be clear
that we need not dress the quark-quark-gluon vertex, as such corrections are
already accounted for by the terms summed in (\ref{Sigma}).}. Thus, we 
obtain \cite{fz2}  
\beq\label{Sigma}
N\Si(\eta; r) = \sum_{k=1}^{\infty} \Gam_{2k} \left[\cg (\eta; r)\right]
^{2k-1}\,,
\eeq
where each of the summed terms is manifestly of order $N$.

The requirement, mentioned above, that the blob corresponding to 
$\Gamma_{2k}$ be inseparable into smaller blobs (a property referred to 
as ``gluon connectedness'' in \cite{bluez}) can be explained with the 
help of the simple example shown in Fig.9.
%%%%%%%%%%%%%%%%%%%%%%
\vspace{24pt}
\par
%\hspace{0.5in} \epsfbox{talk-feynman-rules.eps}
\hspace{0.5in} \epsfbox{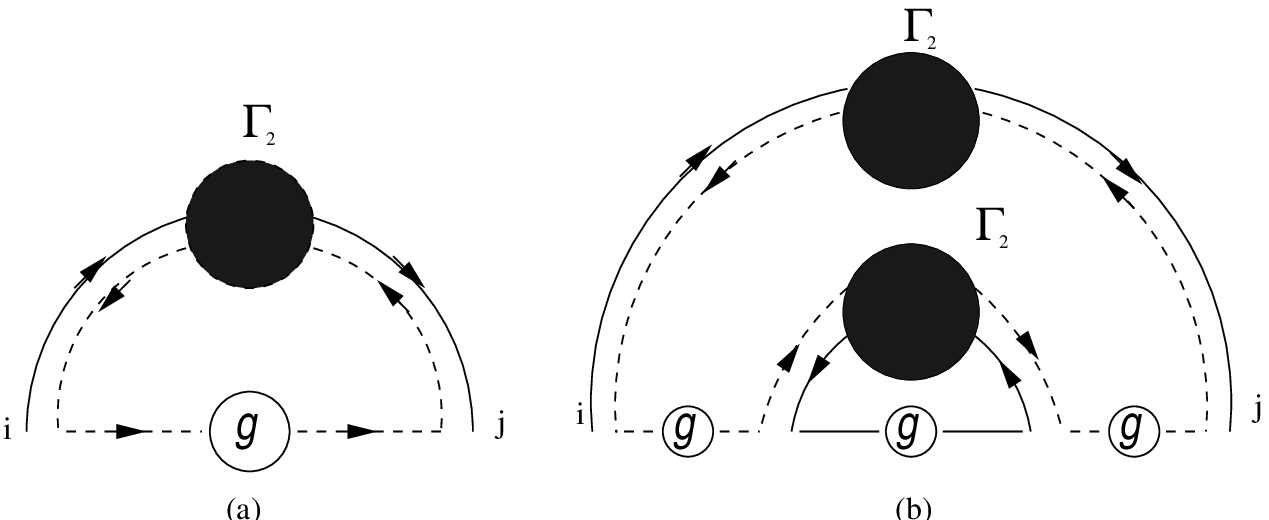}
\par
\baselineskip 12pt
{\footnotesize {\bf Fig.~9:} The contribution of the quadratic cumulant (a)
and that of a separable gluon blob (b). The diagram (b) is evidently already 
accounted for in (a).}
\vspace{24pt}
\par
\baselineskip 20pt   
%%%%%%%%%%%%%%%%%%%%%
Evidently, the diagram in Fig.9(b) is already accounted for by the 
contribution of the second cumulant. It is therefore not included in the 
contribution of the fourth cumulant shown in Fig.8.

If we knew $\Si (\eta;r)$ in terms of $\cg (\eta;r)$, then we can solve for 
$\cg (\eta;r)$ and hence the density of eigenvalues. We are thus faced with 
the problem of determining the $\Gam_{2k}$'s which appears to be a rather 
difficult task\footnote{Except, of course, for the Gaussian ensemble, for 
which only the variance $\Gamma_2\neq 0$, and thus simply $\Si=\Gamma_2 \cg$.}.

It is important to note that the $\Gamma_{2k}$ in (\ref{Sigma}) 
depend only on the probability distribution $P(\phi)$ and not on the 
particular quantity we 
average over. Consider the problem of determining the eigenvalue 
density of the hermitean matrix $\pdgp$. We would study the Green's function 
\beq\label{FF}
F(w) = \langle {1\over N} \trn {1\over w -\pdgp}\rangle \equiv 
\int\limits_0^\infty {\tilde\rho(\si) d\si\over w-\si} \,,
\eeq
where $\tilde\rho(\mu)$ is the averaged eigenvalue density of $\pdgp$. 
We see from (\ref{GH}) that this Green's function is related to 
$\cg (\eta ; r=0)$ simply through
\beq\label{relation} 
\cg (\eta ; r=0) = \eta F( \eta^2 )\,.
\eeq
Also, for $z=0$ we have from (\ref{selfenergy1})
\beq\label{selfenergy2}
\Si (\eta; 0) = \eta - \cg^{-1}(\eta; 0)\,.
\eeq
The crucial observation is that $F(w)$ is already known in the literature on 
chiral and rectangular block random hermitean matrices
for the Gaussian distribution, as well as for non-Gaussian probability 
distributions of the form (\ref{prob}) with an arbitrary polynomial potential 
$V(\pdgp)$)\cite{chiral, chiral1, ambjorn}.
(See also the paper by Feinberg and Zee cited in \cite{rg}.)
In fact, the authors of \cite{chiral1, ambjorn} simply calculated the diagonal 
elements of the propagator $\gmn (\eta;0)$ using Dyson gas techniques (so the 
coefficients $\Gam_{2k}$ are only implicit in these papers.) 

The contents of (\ref{Sigma}), (\ref{relation}) and (\ref{selfenergy2}) may 
thus be summarized as follows: there is a unique function $ a (\xi)$ which 
behaves like $1/\xi$ as $\xi$ tends to infinity, is regular at $\xi=0$, and 
satisfies the equation $\xi-a^{-1}=\sum_{k=1}^{\infty} \Gam_{2k} a^{2k-1}$. 
This function is 
\beq\label{a}
a(\xi) = \xi F(\xi^2) \equiv \cg(\xi; 0)\,. 
\eeq

We, on the other hand, are interested in the opposite case, where $\eta=0, 
r=|z|\neq 0$. Consider the matrix $\gmn (0-i0; z, z^*)$. It follows from 
(\ref{selfenergy}) and (\ref{Sigmamn}) that its $N\times N$ blocks 
are all proportional to the unit matrix ${\bf 1}_N$ (we refer to this as 
``color index democracy"), and we see that 
\beq\label{prop0}
\gmn (0-i0; z, z^*) =\left(\begin{array}{cc} \cg~ & G^*\\{} & {}\\
G~ & \cg~\end{array}\right)\,,
\eeq
where $\cg(0-i0; r)$ is given in (\ref{observe}) and the complex Green's 
function $G(z, z^*)$, the lower left block of $\gmn (0-i0; z, z^*)$, was 
defined \footnote{From this point on, unless 
otherwise stated, our notation will be such that $\cg\equiv\cg(0-i0;r)$, 
$\Si\equiv\Si(0;r)$ and $G\equiv G(z, z^*)$. Note in particular from 
(\ref{observe}) that $\cg$ is then pure imaginary, 
with $\Im \cg\geq 0$.} in (\ref{greens}). Inverting the propagator in 
(\ref{prop0}) and using (\ref{inverseprop}) (with $\eta=0$) and 
(\ref{selfenergy}), we obtain the two equations
\beq\label{zsi1}
z = {G^*\over |G|^2 - \cg^2}
\eeq
and 
\beq\label{zsi11}
\Si = {\cg\over |G|^2 - \cg^2}\,.
\eeq
Note from (\ref{zsi1}) that if $\cg=0$, then $G=1/z$, which corresponds to 
the region outside the domain of the eigenvalue distribution.
In addition, using (\ref{zsi1}) twice, we obtain
\beq\label{zg}
zG = {|G|^2\over |G|^2-\cg^2} = (|G|^2-\cg^2)~r^2\,.
\eeq
Thus, due to the fact that $\cg$ is pure imaginary (see (\ref{observe}))
we conclude that $\gam\equiv zG$ is always real and non-negative.
Alternatively, we can write 
\beq\label{gg}
\cg^2 = {zG~(zG-1)\over r^2}\,,
\eeq
and therefore from the fact that $\cg^2\leq 0$ we conclude that 
$0\leq\gam\leq 1$, in accordance with the monotonicity of $\gam (r)$ and the 
sum-rules $\gam(0)=0$ and $\gam(\infty)=1$ in (\ref{sumrules}).

Let us now define the pure imaginary quantity
\beq\label{xi1}
\xi\equiv {1\over \cg} + \Si = {1\over \cg} + \sum_{k=1}^{\infty} \Gam_{2k} 
\cg^{2k-1}\,,
\eeq
where the last equality follows from (\ref{Sigma}) (with $\eta=0$).
Using (\ref{zsi1}) and (\ref{zsi11}) we thus have simply 
\beq\label{final11}
\xi= {zG\over \cg}\,.
\eeq
It then follows from (\ref{a}) and the statement preceding that equation, and 
from (\ref{final11}) that 
\beq\label{final12}
\xi F(\xi^2) = \cg = {zG\over\xi}\,.
\eeq
We have thus eliminated the $\Gam_{2k}$ !
 
From (\ref{gg}) and from (\ref{final11}) we determine $\xi=\gam/\cg$ to be
\beq\label{xi}
\xi^2 = {\gam \over \gam -1}~r^2\,.
\eeq
Note that $\xi^2\leq 0$, consistent with the bounds $0\leq\gam\leq 1$.
Comparing (\ref{final12}) and (\ref{relation}) we obtain the remarkable 
relation 
\beq\label{remarkablerelation} 
\cg(\xi;0) = \cg(0;r)\,,
\eeq
for $\xi$ and $r$ that are related by (\ref{xi}) (with $\xi$ imaginary.)

We now have the desired equation for $\gam\equiv zG$: substituting (\ref{xi}) 
into (\ref{final12}) we obtain
\beq\label{final}
\gam\left[r^2~F\left({\gam~r^2\over \gam -1}\right) -\gam +1\right] = 0\,.
\eeq
This is the master equation announced in the Abstract. 
Thus, given $F$ we can solve for $\gam(r)$ using this equation.

From (\ref{FF}) we observe that for $w$ large we may expand
\beq\label{expand}
F(w) = {1\over w} + {1\over w^2} f(w)\,,
\eeq
where $f(w) = f_0 + (f_1/w) + \cdots $. Substituting 
$w=\xi^2=\gam r^2/(\gam-1)$ into (\ref{expand}) and letting $\gam$ tend to 
$1$, (\ref{final}) becomes
\beq\label{doubleroot}
(\gam-1)^2\left[{f(w)\over \gam r^2} -1\right]\,,
\eeq
and thus $\gam=1$, which corresponds to the region outside the eigenvalue 
distribution, is always a double root of (\ref{final}).

We will now explain how the Green's function $F(w)$ is obtained in the 
literature\cite{chiral1, ambjorn}. (See also the paper by Feinberg and Zee 
cited in \cite{rg}.) Let $V$ be a polynomial of 
degree $p$. From the definition of $F(w)$ in (\ref{FF}) and its 
analyticity property, we expect $F$, following the arguments of Br\'ezin et al,
to have the form\footnote{Here we assume for simplicity that the eigenvalues 
of $\pdgp$ condense into a single segment $[a,b]$.}
\beq\label{FFF}
F(w)={1\over 2}V'(w)-P(w)\sqrt{(w-a)(w-b)}\,,
\eeq
where 
\beq\label{P}
P(w) = \sum_{k=-1}^{p-2} c_k~w^k\,.
\eeq
The constants $0 \leq a < b$ and $c_k$ are then determined completely
by the requirement that $F(w)\rightarrow {1\over w}$ as $w$ tends to infinity,
and by the condition that $F(w)$ has at most an integrable 
singularity as $w\rightarrow 0$. (Thus, if $a>0$, inevitably 
$c_{-1} = 0$. However, if $a=0$, then $c_{-1}$ will be determined by the 
first condition.)

\section{An Example: the Quartic Ensemble}
Having determined $F(w)$ in this way, we substitute it into (\ref{final}) and 
find $G(z, z^*)$. Equation (\ref{final}) is an algebraic equation for 
$\gam(r)$ and thus may have several $r$ dependent solutions. In constructing 
the actual $\gam(r)$ one may have to match these solutions smoothly into a 
single function which increases monotonically from $\gam(0)=0$ to 
$\gam(\infty)=1$ (recall (\ref{sumrules})). (An explicit non-trivial example 
of such a sewing procedure is the construction of $\gam(r)$ in the disk phase 
of the quartic ensemble\cite{fz2}.) We can thus calculate the density of eigenvalues 
explicitly for an arbitrary $V$. 

The Gaussian case with $V=\pdgp$ provides
the simplest example. Here we have $2\xi F(\xi^2) = \xi - \sqrt{\xi^2 -4}$, 
whence the roots of (\ref{final}) are $\gam=0, 1$ and $r^2$. We note that 
$\gam=0$ is unphysical, $\gam=r^2$ ({\em i.e.,} $G=z^*$) corresponds to 
Ginibre's disk \cite{ginibre}, and $\gam=1$ is the solution outside the disk.

As an explicit non-trivial example, the quartic ensemble with 
$V(\pdgp)=2m^2\pdgp + g (\pdgp)^2$ was studied in detail in \cite{fz2}. 
As should perhaps be expected in advance, the following behavior in the 
parameter space $m^2, g>0$ was found: for $m^2$ positive, the eigenvalue 
distribution was disk-like (and non-uniform), generalizing Ginibre's work, 
but as $m^2\equiv -\mu^2$ was made more and more negative, a phase transition 
at the critical value $\mu_c^2 = \sqrt{2g}$
occurred, after which the disk fragmented into an annulus. 
The density of eigenvalues was calculated in \cite{fz2} in 
detail. In particular, the radii of the disk and annulus were calculated
explicitly in terms of the couplings in $V$. An interesting behavior
of the density of eigenvalues was observed through the disk-annulus phase
transition. As the transition was approached from the disk side (e.g., by 
lowering $m^2$ towards $-\mu_c^2$ holding $g$ fixed), a central region of 
{\em finite} radius in the disk got progressively depleted. At the transition
depletion was complete and an annulus of finite critical radius was formed, 
with vanishing eigenvalue density at the inner radius of the annulus. 
As the couplings were tuned deeper into the annular phase, a finite jump in 
the density of eigenvalues at the inner radius appeared. The interested reader
could find all relevant formulas in Section 5 of \cite{fz2} and Section 4 of 
\cite{fsz}. In what follows we shall compare these analytical large-$N$ 
results, following \cite{fsz}, against numerical simulations.

In Fig.10 we display the numerical results for $\rho_{disk} (r)$ for 
128x128 dimensional matrices, and compare them to the analytical large-$N$ 
result (5.8) of \cite{fz2} (or (4.4) of \cite{fsz}). (As a trivial check of 
our numerical code, we also included in this figure the results for the 
gaussian (Ginibre) ensemble.)
Evidently, the agreement between the numerical and the analytical 
results is good. Note the finite-$N$ effects near the edge of the disk. 
%%%%%%%%%%%%%%%%%%%%%%
\vspace{-34pt}
\par
\hspace{0.5in} \epsfbox{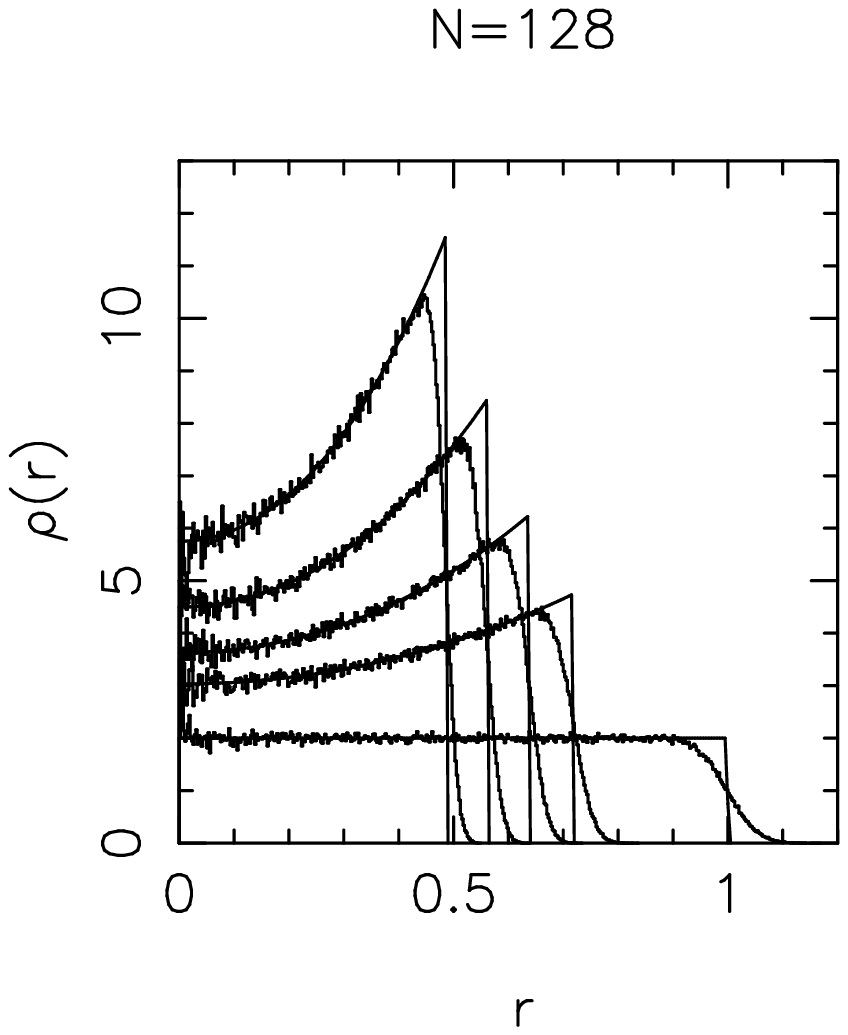}
\par
\baselineskip 12pt
\vspace{-24pt}
{\footnotesize {\bf Fig.~10:} Comparison between Monte-Carlo measurements of 
the density of eigenvalues $\rho (r)$ of matrices $\phi$ of size 128x128, 
taken from the quartic ensemble $V(\pdgp) = 2m^2\pdgp + g(\pdgp)^2 $ with 
$m^2=0.5$ (disk phase) and for $g = 0, 0.5, 1, 2, 4$ ($g$ increases from 
bottom to top), compared to the analytical results of \cite{fz2} 
(solid lines). At $g=0$ we obtain Ginibre's Gaussian ensemble with 
$V=\pdgp$, with its unit disk of eigenvalues. (This figure is taken from 
\cite{fsz}.)}
\vspace{24pt}
\par
\baselineskip 20pt   
%%%%%%%%%%%%%%%%%%%%%
In Figs11.(a)-(c) we display our numerical results for 
$\rho_{annulus} (r)$ for matrices of various sizes, 
and compare them to the analytical large-$N$ result (5.18) of \cite{fz2} (or 
(4.13) of \cite{fsz}). In these figures we hold $-m^2=\mu^2=0.5$ fixed, and 
increase $g$ from $0.025$ to $0.1$. (Here we have $\mu^2= 0.5 = 
\mu_c^2/2\sqrt{2g}$. Thus increasing $g$ as indicated in the text brings us 
closer to the disk-annulus phase transition.)
%%%%%%%%%%%%%%%%%%%%%%
\vspace{-14pt}
\par
\hspace{0.5in} \epsfbox{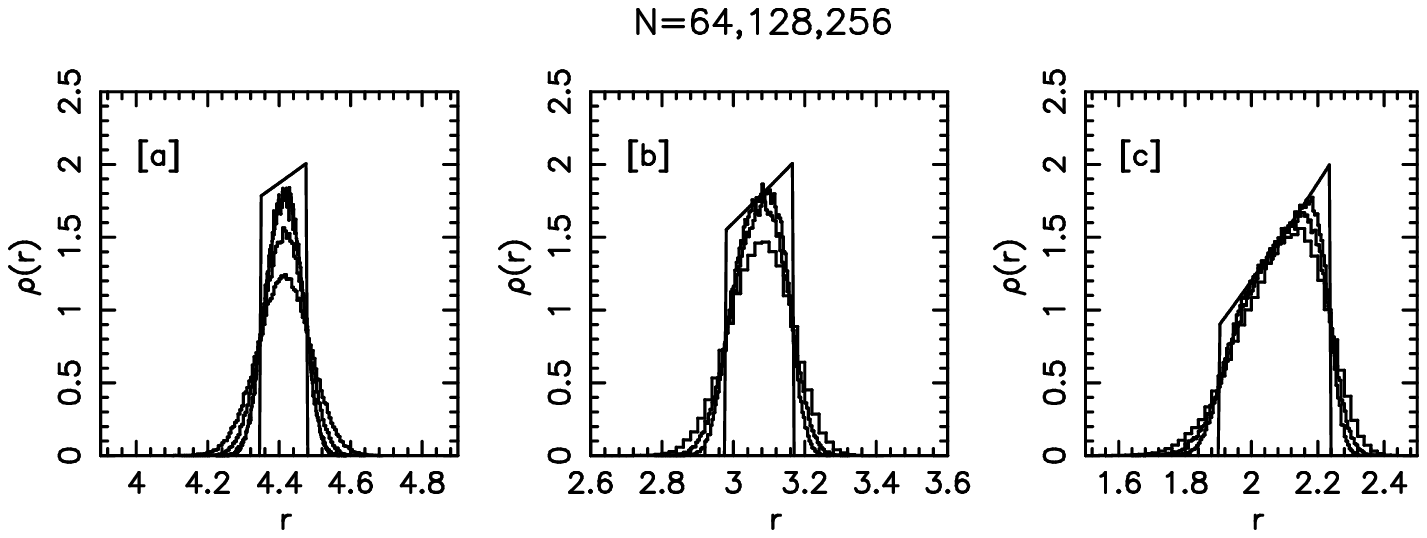}
\par
\baselineskip 12pt
{\footnotesize {\bf Fig.~11:}
Results of Monte-Carlo measurements of the density of 
eigenvalues $\rho (r)$ of matrices $\phi$ of sizes corresponding to 
$N=64, 128$ and $256$, taken from the quartic ensemble with 
$m^2=-\mu^2 =-0.5$ (annular phase) for various values of the quartic 
coupling: $g=0.025$ in [a], $g=0.05$ in [b] and $g=0.1$ in [c]. These are 
compared to the analytical results of 
\cite{fz2} (solid lines). As $N$ increases, the numerical results 
converge monotonically to the analytical results. (This figure is taken
from \cite{fsz}.)}
\vspace{14pt}
\par
\baselineskip 20pt   
%%%%%%%%%%%%%%%%%%%%%
In \cite{fsz} we have also studied numerically the disk-annulus phase 
transition. We have measured the density of eigenvalues 
$\rho (r)$ of matrices $\phi$ of size 128x128, taken from the 
the quartic ensemble with $\mu^2=0.5$ and for 
$g = 0.025, 0.05, 0.1, 0.125, 0.15$ and $0.175$. The results are 
displayed on Figure 12. 

For these values of $g$, we start in the annular phase 
at the lowest value of $g$. For our set of parameters we have $\mu^2= 0.5 = 
\mu_c^2/2\sqrt{2g}$. Thus, increasing $g$ (while keeping $\mu^2$ fixed at 
$0.5$) brings us closer to the disk-annulus phase transition, which occurs 
(at large $N$) at $g_c=0.125$. Increasing $g$ 
beyond that, puts us into the disk phase.  

The first three profiles on the right in Figure 12 belong to the annular 
phase. As $g$ increases towards the transition point at $g_c=0.125$, these 
three graphs exhibit decrease of the inner radius of the annulus, in 
accordance with the analytical results.

The critical density profile, corresponding to 
$g_c=0.125$, is the fourth profile (from the right). For our choice of 
parameters, the theoretical boundary radii of the critical annulus, i.e., at 
$g=0.125$, are $R_{in}^{crit} = 1/\mu_c = \sqrt{2}$ and 
$R_{out}^{crit}=\sqrt{2}/\mu_c=2$. These boundary values fit nicely with 
the features of the critical profile in Figure 12.

Finally, the last two profiles in Figure 12 have pronounced tails extending 
to $r=0$ and thus belong to the disk phase.
%%%%%%%%%%%%%%%%%%%%%%
\vspace{-44pt}
\par
\hspace{0.5in} \epsfbox{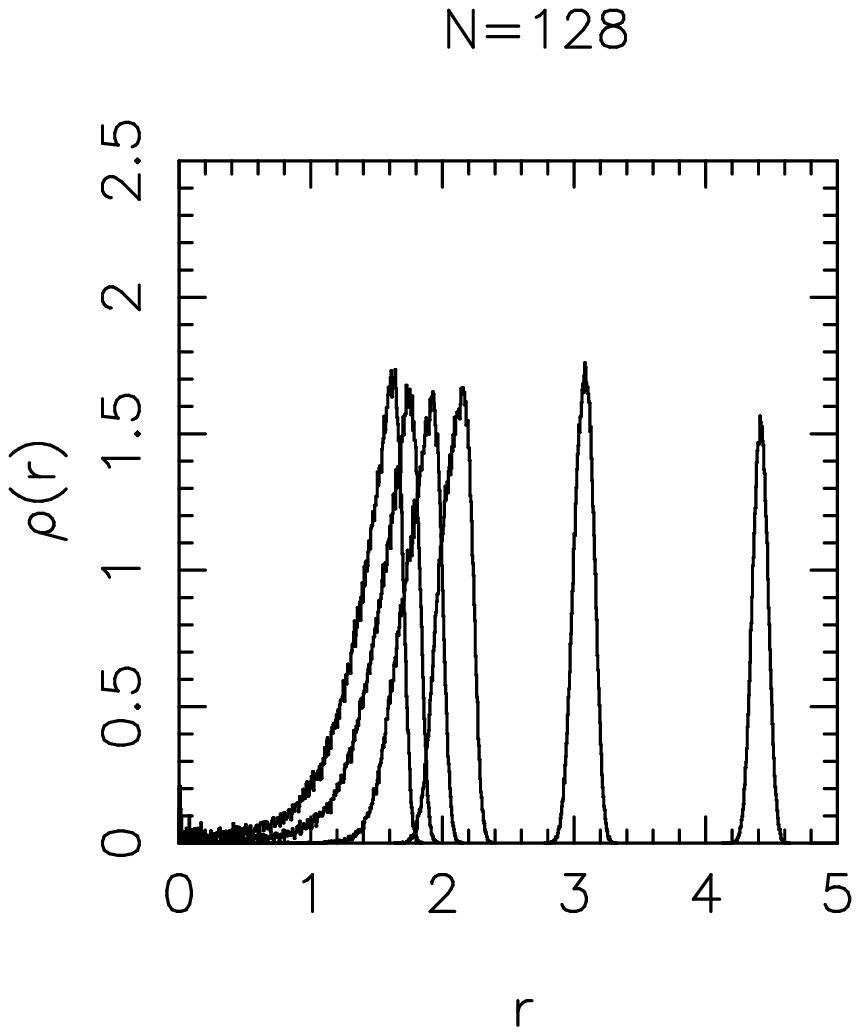}
\par
\baselineskip 12pt
{\footnotesize {\bf Fig.~12:}
Monte-Carlo measurements of the density of eigenvalues 
$\rho (r)$ of matrices $\phi$ of size 128x128, taken from the 
the quartic ensemble with $\mu^2=0.5$ and for 
$g = 0.025, 0.05, 0.1, 0.125, 0.15$ and $0.175$ ($g$ increases from 
right to left). The first three profiles on the right (corresponding to the 
three lowest values of $g$) evidently belong to the annular phase. The fourth 
density profile from the right is the critical one 
(corresponding to $g_c=0.125$). Finally, The last two profiles 
(which correspond to the two higher values of $g$) belong to the 
disk configuration. (This figure is taken from \cite{fsz}.)}
\vspace{14pt}
\par
\baselineskip 20pt   
%%%%%%%%%%%%%%%%%%%%%

\section{Boundaries and Shape Universality of the Eigenvalue Distribution:
the Single Ring Theorem}

A remarkable property of (\ref{final}) is that it has only two
$r$-independent solutions: $\gam=0$ and $\gam =1$ \cite{fz2}.
Since the actual $\gam(r)$ increases monotonically from 
$\gam(0)=0$ to $\gam(\infty)=1$, without any plateaus in between, 
we immediately conclude from 
this observation that there can be no more than a single void 
in the eigenvalue distribution. Thus, in the class of models  governed by 
$P(\phi) = {1\over Z} e^{-N\rmtr V(\phi^\dgg\phi)}$
(Eq. (\ref{prob})), the shape of the eigenvalue 
distribution is either a disk or an annulus, whatever 
polynomial the potential $V(\pdgp)$ is. This result is the ``Single Ring 
Theorem" of \cite{fz2}.

The ``Single Ring Theorem" may appear counter-intuitive at 
first sight. Indeed, consider a potential $V(\pdgp)$ with 
several wells or minima. For deep enough wells, we expect the 
eigenvalues of $\pdgp$ to ``fall into 
the wells". Thus, one might suppose that the eigenvalue 
distribution of $\phi$ to be bounded by a set of concentric 
circles of radii $0\leq r_1 < r_2 < \cdots < r_{n_{\rm max}}$, 
separating annular regions on which $\rho(r)>0$ from voids 
(annuli in which $\rho(r)=0$.) A priori, it is natural to 
assume that the maximal number of such circular boundaries 
should grow with the degree of $V$, because $V$ may then have 
many deep minima. Remarkably, however, according to the 
``Single Ring Theorem" the number of these 
boundaries is two at the most.

To reconcile this conclusion 
with the a priori expectation just mentioned, note that while the eigenvalues 
of the hermitean matrix $\pdgp$ may split into several disjoint segments 
along the positive real axis, this does not necessarily constrain the 
eigenvalues of $\phi$ itself to condense into annuli. Indeed, the hermitean 
matrix $\pdgp$ can always be diagonalized $\pdgp=U^\dgg\La^2 U$ by a unitary 
matrix $U$, with $\La^2={\rm diag}(\la_1^2, \la_2^2,\cdots,\la_N^2),$ where 
the $\la_i$ are all real. This implies that $\phi=V^\dgg\La U$, with $V$ a 
unitary matrix as well. Thus, the complex eigenvalues of $\phi$ are given by 
the roots of ${\rm det} (z-\La W)=0$, with $W=UV^\dgg$. Evidently, as $W$ 
ranges over $U(N)$ (which is what we expect to happen in the generic case), 
the eigenvalues of $\La W$ could be smeared (in the sense that they would 
not span narrow annuli around the circles $|z|=|\la_i|$.) 

The last argument in favor of the ``Single Ring Theorem" 
clearly breaks down when $W$ fails to range over $U(N)$, which occurs 
when the unitary matrices $U$ and $V$ are correlated. For example, $\phi$   
may be such that $W=UV^\dgg$ is block diagonal, with the upper diagonal block 
being a $K\times K$ unitary diagonal matrix ${\rm diag}(e^{i\om_1}, 
\cdots , e^{i\om_K})$ (and with $K$ a finite fraction of $N$). In the extreme 
case $K=N$, in which $W$ is completely diagonal, 
$W\equiv e^{i\om} = {\rm diag}(e^{i\om_1}, \cdots , e^{i\om_N})$, we see that 
$\phi = U^\dgg e^{i\om}\Lambda U$ is a {\em normal} matrix (that is,
$[\phi, \phi^\dgg]=0$) with eigenvalues ${\rm diag}(e^{i\om_1}\la_1, \cdots ,
e^{i\om_N}\la_N ).$ Thus, normal matrices, or partially normal matrices 
(i.e., the case $K<N$), evade the ``Single Ring'' theorem: if the 
first $K$ eigenvalues $\la_1^2, \la_2^2,\cdots,\la_K^2$ of 
$\pdgp$ split into several disjoint segments along the positive real axis, 
the corresponding eigenvalues of $\phi$ will split into concentric annuli in 
the complex plane obtained by revolving those $\la$-segments. Normal, 
or partially normal matrices are, of course, extremely rare in the 
ensembles of non-hermitean matrices studied in this paper, and do not 
affect the ``Single Ring'' behavior of the bulk of matrices in the ensemble.

Let us sketch now the proof of the Single Ring Theorem, following \cite{fz2}. 
Thus, let us assume for the moment that the domain of eigenvalues has 
$n$  boundaries.  It is easy to see from (\ref{circular}), the defining 
equation of $\gam(r)$, that in the annular void $r_k<r<r_{k+1}$ in the 
eigenvalue distribution, $\gam$ is a constant which is equal to the 
fraction of eigenvalues contained inside the 
disk $r\leq r_k$. Thus, the equation to determine $\gam$ (Eq. (\ref{final})
which we repeat here for convenience)
\beq\label{final6}
\gam\left[r^2~F\left({\gam~r^2\over \gam -1}\right) -\gam +1\right] = 0
\eeq
must have a series of monotonically increasing constant solutions 
$\gam_1<\gam_2<\cdots \leq 1$, which correspond to the various voids.

In particular, from (\ref{sumrules}) we have $\gam =1$ for 
$r>r_{n_{\rm max}}$, namely, $G=1/z$. Thus, $\gam=1$ must be a solution of 
(\ref{final6}), the maximal allowed constant solution, as we already saw 
following (\ref{doubleroot}). 
Also, if $r_1>0$, namely, if there is a hole at the center of the eigenvalue 
distribution, then for $r\leq r_1$ we must have $\gam =0$ (independently of 
$r$), which is obviously a solution of (\ref{final6}). (From the paragraph 
right below (\ref{FFF}) it is clear that $w F(w)\rightarrow 0$ at $w=0$. Thus
$\gam =0$ is indeed a solution. On the other hand, if $r_1=0$, so that the 
eigenvalue distribution includes the origin, an $r$ independent solution 
$\gam=0$ is of course only a spurious solution which should be discarded.)

Assume now that $\gam=\gam_0$ is an $r$ independent solution of (\ref{final6}).
Taking the derivative of (\ref{final6}) with respect to $r^2$ at 
$\gam\equiv\gam_0$ we obtain 
\beq\label{gam0}
\gam_0\left[ F(\xi^2) + \xi^2{dF\over d\xi^2}\right]_{|_{\xi^2=
{\gam_0 \over \gam_0 -1}r^2}} =0\,,
\eeq
which is the condition for the existence of an $r$ independent solution 
$\gam_0$. Thus, there are two possibilities: either $\gam_0=0$, which we 
already encountered, or $ F(\xi^2) + \xi^2{dF\over d\xi^2}=0$. This equation
immediately yields
\beq\label{Fxi}
F(\xi^2)={1\over \xi^2}\,,
\eeq
where the integration constant is fixed by the asymptotic behavior of 
$F(\xi^2)$ as $\xi\rightarrow\infty$. But for a generic $V(\pdgp)$, 
$F(w)$ is given by (Eq. (\ref{FF}))
\beqast
F(w) = \langle {1\over N} \trn {1\over w -\pdgp}\rangle \equiv 
\int\limits_0^\infty {\tilde\rho(\si) d\si\over w-\si} \,,
\eeqast
with (\ref{Fxi}) being the asymptotic behavior of $F(w)$ as
$w=\xi^2\rightarrow\infty$. We thus conclude that $\xi^2\rightarrow\infty$
in (\ref{Fxi}), namely, that $\gam_0=1$. Thus, to summarize, the only 
possible $r$ independent solutions of (\ref{final}) are $\gam=0$ and 
$\gam=1$, which we already discussed. Since no other $r$ independent 
solutions arise, there can be no more than a single void in the eigenvalue 
distribution, whatever polynomial the potential $V(\pdgp)$ is. 
The shape of the eigenvalue distribution is thus either a disk or an annulus.

Interestingly enough, we can arrive at the same conclusion by invoking other 
general aspects of the method of hermitization, and thus providing a nice 
self consistency check of our formalism. Recall at this point that the 
boundaries of the eigenvalue distribution are given by (\ref{boundary}), 
namely, the zeros of $\cg$. But $\cg$ is given by (Eq. (\ref{gg}))
\beqast
\cg^2 = {\gam (\gam-1)\over r^2}\,,
\eeqast
from which we see that at the boundaries $\gam$ can take on only two values : 
zero and one. Since $\gam$ is a constant in the void, by continuity, these 
are the only possible values of $\gam$ inside any of the voids. Therefore, 
there may be two circular boundaries at most. In other words, as far as the 
eigenvalue density of $\phi$ is concerned, an ensemble of the form (\ref{prob})
may have two phases at most, as we concluded earlier.

In addition to $\gam = 1$ (and possibly $\gam = 0$), (\ref{final6}) 
must have other roots which do depend on $r$. Among all these other roots, 
we expect to find a unique root $\gam(r)$, which is a positive monotonically 
increasing function of $r$,  that matches continuously at the boundaries 
$r_k\quad (k\leq 2)$ to the constant roots of (\ref{final6}). The actual 
$zG(z, z^*)$ of the ensemble (\ref{prob}) is therefore a continuous 
monotonically increasing function of $r$, as required by (\ref{circular}).
If the eigenvalue distribution is annular with boundaries $0\leq r_1<r_2$ 
(the case $r_1 =0$ corresponding to the full disk), it vanishes for $0\leq 
r\leq r_1$, rises monotonically from zero to one on $r_1\leq r\leq r_2$, and 
equals to one for $r\geq r_2$.

The sextic potential $V(\phi^\dgg\phi) = m^2 \pdgp +{\lambda\over 2} (\pdgp)^2
+ {g\over 3} (\pdgp)^3$
is the potential of lowest degree in (\ref{prob}) for which the eigenvalues 
of $\pdgp$ may split into more then a single segment. Therefore, it is the
simplest case in which the Single Ring theorem has non-trivial consequences. 
In fact, it is easy to see that there can be at most two eigenvalue segments 
in the spectrum of 
$\pdgp$, in which case, one segment touches the origin. In \cite{fsz} we used 
Monte-Carlo simulations to generate an ensemble of random matrices 
(of size $32\times 32$) corresponding to (\ref{prob}) with the sextic 
potential. The coefficients of the latter were tuned to the phase
where the spectrum of $\pdgp$ splits into two segments. We verified numerically
that the eigenvalue distribution of $\phi$ in this phase was always a disk, and
not a disk surrounded by a concentric annulus.

\section{Universal Features of the Disk and Annular Phases and the 
Transition Between Them}

According to the ``Single Ring" theorem, the 
eigenvalue distribution of $\phi$ is either a disk or an annulus. 
The behavior of $F(w)$ as $w\sim 0$ turns out to be an indicator 
as to which phase of the two the system is in \cite{fsz}, as I now show:

{\em A. Disk Phase:}~~~In the disk phase we expect that $\rho(0)>0$, as in 
Ginibre's case. 
Thus, from (\ref{rhocirc}) $\rho(r) = (1/r) (d\gam/dr) \equiv 2 
(d\gam/dr^2)$ and from the first sum rule $\gam(0)=0$ in (\ref{sumrules}) we 
conclude that 
\beq\label{near0}
\gam(r)\sim {1\over 2} \rho(0) r^2
\eeq
near $r=0$. 
Therefore, for $r$ small, 
(\ref{final}) yields 
\beq\label{near00}
F\left(-{\rho(0) r^4\over 2} + \cdots \right)\sim - {1\over r^2}\,,
\eeq 
namely, $F(w)\sim 1/\sqrt{w}$ for $w\sim 0$, as we could
have anticipated from Ginibre's case. This means
that in the disk phase we must set $a=0$ in (\ref{FFF}). Consequently, in the 
disk phase $c_{-1}$ does not vanish. We can do even better: paying 
attention to the coefficients in (\ref{FFF}) and (\ref{P}) (with $a=0$) we 
immediately obtain from (\ref{near00}) that 
\beq\label{rho0}
c_{-1} = \sqrt{\rho(0)\over 2 b}\,.
\eeq

{\em B. Annular Phase:} In the annular phase $\gam(r)$ must clearly  
vanish identically in the inner void of the annulus.  Thus, 
(\ref{final}) implies that $F(w)$ cannot have a pole at $w=0$, and therefore 
from 
(\ref{FFF}) we must have $c_{-1}\sqrt{ab}=0$. Thus, the annulus must 
arise for $c_{-1} = 0$ (the other possible solution $a=0, c_{-1}\neq
0$ leads to a disk configuration with $\gam=0$ only at $r=0$, as we just 
discussed.)

Thus, to summarize, in the disk phase $F$ has the form 
\beq\label{Fdisk}
\!\!\!\!\!\!\!\!\!\!\!\!\!\!\!\!\!\!
 F_{disk}(w)={1\over 2}V'(w)-\left(\sqrt{{\rho(0)\over 2b}}
\,w^{-1} + c_0 + c_1~w +\cdots + c_{p-2}~w^{p-2}\right)\sqrt{w(w-b)}\,,
\eeq
while in the annular phase it has the form
\beq\label{Fannulus}
\!\!\!\!\!\!\!\!\!\!\!\!\!\!\!\!\!\!
F_{annulus}(w)={1\over 2}V'(w)-\left(c_0 + c_1 ~w +\cdots 
+c_{p-2}~w^{p-2}\right)\sqrt{(w-a)(w-b)}\,. \eeq
Having determined $F(w)$ in this way, i.e., having determined the various 
unknown parameters 
in (\ref{Fdisk}) or in (\ref{Fannulus}), we substitute it into (\ref{final}) 
and find $G(z, z^*)$. We can thus calculate the density of eigenvalues 
$\rho(r)$ explicitly for an arbitrary $V$. 

{\em C. The Disk-Annulus Phase Transition:}
We now turn to the disk-annulus phase transition. An important feature of 
this transition is that $F(w)$ is continuous
through it. To see this we argue as follows: By tuning the couplings in $V$, 
we can induce a phase transition from the 
disk phase into the annular phase,
or vice versa. Note, of course, that we can parametrize any point in the disk 
phase either by the set of couplings in
$V$ or by the set of parameters $\{c_{-1}, c_0,  \cdots c_{p-2}; b\}$ in 
(\ref{Fdisk}). The ``coordinate transformation" between
these two sets of parameters is encoded in the asymptotic behavior of $F(w)$. 
Similarly, we can parametrize any point in the annular phase either by the 
set of couplings in $V$ or by the set of parameters $\{c_0, c_1 \cdots 
c_{p-2}; a, b\}$ in (\ref{Fannulus}). Due to the one-to-one relation (in a 
given phase, once we have established it is the stable one) between the 
couplings in $V$ and the parameters in $F(w) - {1\over 2} V'(w)$ (namely, the 
$c_n$'s and the locations of the branch points of $F(w)$), we can 
describe the disk-annulus transition in terms of the latter parameters 
(instead of the couplings in $V$). Clearly, the 
transition point is reached from the disk phase when $\rho(0) = 0$, that is, 
when $c_{-1}$ in (\ref{Fdisk}) vanishes:
\beq\label{cminus1crit}
c_{-1}^{crit} = 0\,.
\eeq
Similarly, the transition point is reached from the annular phase when the 
lower branch point $a$ in (\ref{Fannulus}) vanishes. 
Thus, e.g., in a transition from the disk phase into the annular phase, 
$F_{disk}(w)$ in (\ref{Fdisk}) would cross-over 
continuously into $F_{annulus}(w)$ in (\ref{Fannulus}) through a critical form 
\beq\label{Fcrit}
F_{crit}(w)={1\over 2}V_{crit}'(w)-\left(c^{crit}_0 + c^{crit}_1 ~w +\cdots 
+c^{crit}_{p-2}~w^{p-2}\right)\sqrt{w(w-b^{crit})}\,.\eeq

The continuity of $F(w)$ through the transition was demonstrated explicitly 
in \cite{fz2} for the quartic ensemble $V(\pdgp) = 2m^2 \pdgp + g(\pdgp)^2$. 

This discussion obviously generalizes to cases when $F(w)$
has multiple cuts, which correspond to condensation of the
eigenvalues of $\pdgp$ into many segments. If $w=0$ is a branch
point of $F(w)$, that is, if the lowest cut extends to the
origin, we are in the disk phase,
\beq\label{Fdisk1}
\!\!\!\!\!\!\!\!\!\!\!\!\!\!\!\!\!\!
F_{disk}(w)={1\over 2}V'(w)-\left(c_{-1}w^{-1} + c_0 + c_1 ~w 
+\cdots + c_{p-2}~w^{p-2}\right)\sqrt{w(w-b_1)\cdots (w-b_n)}\,,
\eeq
with $0< b_1 < \cdots < b_n $. The relation (\ref{rho0})
then generalizes to 
\beq\label{rho00}
c_{-1} = \sqrt{{\rho(0) \over 2 (-1)^{n+1} \prod_{k =1}^n b_k}}\,.
\eeq
Since $c_{-1}$ must be real we conclude that such a configuration
exists only for $n$ odd. 

If the lowest branch 
point in $F(w)$ is positive, we are in the annular phase with 
\beq\label{Fannulus1}
\!\!\!\!\!\!\!\!\!\!\!\!\!\!\!\!\!\!
F_{annulus}(w)={1\over 2}V'(w)-\left(c_0 + c_1 ~w +\cdots 
+c_{p-2}~w^{p-2}\right)\sqrt{(w-a)(w-b_1)\cdots (w-b_n)}\,. \eeq
The phase transition would occur when the couplings in $V(\pdgp)$ are tuned 
such that $F_{disk}(w)$ and $F_{annulus}(w)$ match continuously, as 
was described in the previous paragraph.

\section{Boundaries and Boundary Values}
Remarkably, with a minimal amount of effort, and based on the mere
definition of $F(w)$ (Eq. (\ref{FF})), we were able to derive in \cite{fsz}
simple expressions for the location of the 
boundaries of the eigenvalue distribution and also for the 
boundary values of $\rho(r)$ in terms of the moments of $\tilde\rho(\si)$, 
which, I remind the reader, 
is the density of eigenvalues for a hermitean matrix problem. 

To this end it is useful to rewrite our master formula (\ref{final}) for
$\gam (r)$ as 
\beq\label{fin}
w F(w) = \gam
\eeq
with 
\beq\label{w}
w = {\gam r^2\over \gam -1}\,.
\eeq
We start with the outer edge $r=R_{out}$ (either in the disk phase or
in the annular phase.) Near the outer edge $\gam\rightarrow 1-$, and thus 
$w\rightarrow -\infty$. We therefore expand $F(w)$ in (\ref{fin}) in powers 
of $1/w$. As can be seen from (\ref{FF}), the coefficient of 
$w^{-k-1}$ in this expansion is the $k$-th moment 
\beq\label{moments}
<\si^k> = \int\limits_0^\infty \tilde\rho(\si) \si^k d\si
\eeq
of $\tilde\rho(\si)$ (which is of course normalized to 1.)
For the class of models we are interested in here, all the 
moments $<\si^k>,~ k\geq 0$ are finite. 
From the expanded form of (\ref{fin}) one can obtain, by a straightforward
calculation \cite{fsz}, that 
\beq\label{rout}
R_{out}^2 = <\si> 
\eeq
and 
\beq\label{rhoout}
\rho (R_{out})  = {2R_{out}^2\over <\si^2> - <\si>^2}\,.
\eeq
Thus, $R_{out}^2$ is simply the first moment of $\tilde\rho(\si)$, 
and the density $\rho(R_{out})$ is inversely proportional to the variance
of $\si$.

For the $\tilde\rho(\si)$ under consideration here, $<\si^2>$, 
and consequently $\rho(R_{out})$, are always finite. Outside the
boundary $\rho(r)$ vanishes identically, of course, and thus, $\rho(r)$ 
always ``falls off a cliff" at the boundary, for all probability 
distributions of the form (\ref{prob}) with $V$ polynomial. 
It would be thus interesting to study circularly invariant matrix
ensembles $P(\pdgp)$ such that the eigenvalue distribution 
$\tilde\rho(\si)$ of $\pdgp$ has a finite $<\si>$ but an infinite 
$<\si^2>$. Then $\rho(R_{out})$ would vanish. This would naturally raise 
the question whether in such situations, $\rho(r)$ behaves universally 
near the edge (that is, if near the edge it vanishes like $(R_{out} - 
r)^\epsilon$ with $\epsilon$ being some universal exponent).

We now turn to the annular phase, and focus on the inner edge 
$r=R_{in}$ of the annulus. According to the discussion in the previous 
section (see Eq. (\ref{Fannulus}) and the discussion above it), $a>0$ in 
(\ref{FFF}), and thus $F(w)$ is analytic in the domain $|w|<a$. 
We therefore expand $F(w)$ in (\ref{fin}) in powers 
of $w$. As can be seen from (\ref{FF}), the coefficient of 
$w^{k-1}$ in this expansion is the negative moment 
\beq\label{negmoments}
\langle{1\over \si^k}\rangle = \int\limits_0^\infty \tilde\rho_{annulus}(\si) 
{1\over \si^k} d\si
\eeq
of $\tilde\rho_{annulus}(\si).$
From (\ref{FFF}) (or (\ref{Fannulus}) ) it is possible to see that 
all these negative moments of $\tilde\rho_{annulus}(\si)$ exist.
From this expanded form of (\ref{fin}) one can obtain, again, 
by a straightforward calculation \cite{fsz}, that 
\beq\label{rin}
{1\over R^2_{in}} = \langle {1\over \sigma}\rangle
\eeq
and 
\beq\label{rhoin}
\rho (R_{in})  = {2R_{in}^{-6}\over <\si^{-2}> - <\si^{-1}>^2}\,.
\eeq
Thus, $R^{-2}_{in}$ is simply the $\si^{-1}$ moment of $\tilde\rho_{annulus}
(\si)$, and the density $\rho(R_{out})$ is inversely proportional to the 
variance of $\si^{-1}\,.$
Since $<1/\si^k>\,,~k=1,2$ are finite, $\rho_{annulus}(\si)$ jumps from zero 
(in the inner void of the annulus) to a finite value at the inner edge 
$R_{in}$. It can be shown \cite{fsz}, however, that when $a\rightarrow 0$, 
that is, in the annular 
to disk transition, $<1/\si>$ remains finite, but $<1/\si^2>$ diverges 
like $1/\sqrt{a}$. Thus, from (\ref{rin}) we see that $R_{inner} (a=0)$, the 
{\em critical} inner radius, is finite. The annulus starts up with a finite 
inner radius. Also, in this limit, we see from (\ref{rhoin}) that 
$\rho(R_{in})$ vanishes like $\sqrt{a}$. As we approach the annulus-disk 
transition, the discontinuity in $\rho(r)$ at the (finite) inner edge 
disappears.

We saw at the end of the previous section (see Eqs. (\ref{Fdisk})
-(\ref{Fcrit})) that $F(w)$ is continuous through the disk-annulus phase 
transition. Thus, our master formula $wF(w)=\gam $ to determine $\gam(r)$ 
(Eq. (\ref{fin})) is also continuous through the transition. Consequently, 
$\rho(r) = (1/r) (d\gam/dr)$ must remain continuous through the disk-annulus 
transition, and has (at the transition) the universal behavior described in 
the previous paragraph.

\textbf{Acknowledgements} It is a great pleasure to thank A. Zee and 
R. Scalettar for a fruitful  and exciting collaboration. I also thank 
the organizers of of the workshop for inviting me to present this work and 
for the very kind hospitality at the University of Stellenbosch during the 
conference.


\section*{References}

\begin{thebibliography}{99}                                                  
                                            
\bibitem{fz1} J. Feinberg and A. Zee, Nucl. Phys. {\bf B504}, 579 (1997).

\bibitem{fz2} J. Feinberg and A. Zee, Nucl. Phys. {\bf B501}, 643 (1997).

\bibitem{fsz} J. Feinberg, R. Scalettar and  A. Zee, J. Math. Phys. {\bf 42}, 
5718 (2001).

\bibitem{qcd} M. A. ~Stephanov, Phys. ~Rev. ~Lett. {\bf 76 } (1996) 
4472. ~M.A. Halasz, A.D. Jackson, R.E. Shrock, M.A. Stephanov and 
J.J.M. Verbaarschot, Phys. ~Rev. {\bf D 58} (1998) 096007. 
~J. J. M. Verbaarschot and T. Wettig, {\sl Random Matrix Theory 
and Chiral Symmetry in QCD}, ~Ann.~Rev.~Nucl.~Part.~Sci. {\bf 50} (2000) 343.
\newline
~J. J. M. Verbaarschot, {\sl QCD, Chiral Random Matrix Theory and 
Integrability}, Lectures given at the Les Houches Summer School on 
Applications of Random Matrices in Physics, Les Houches, France, 
June 2004,  hep-th/0502029.
\newline M. A. Nowak, {\sl Lectures on Chiral Disorder in QCD}, Lectures 
given at Cargese Summer School on QCD Perspectives on Hot and Dense Matter, 
Cargese, France, August 2001, hep-ph/0112296.
~R. A. Janik, M. A. Nowak, G. Papp and I. Zahed, {\sl Chiral Random Matrix 
Models in QCD}, ~Acta ~Phys.Polon. {\bf B29} (1998) 3957;~ 
{\sl Various Shades of Blue's Functions},~Acta~Phys.~Polon. {\bf B 28} 
(1997) 2949.  

\bibitem{zahed} R.A. Janik, M.A. Nowak, G. Papp, J. Wambach and I. Zahed, 
Phys. Rev. E {\bf 55}, 4100 (1997); R.A. Janik, M.A.
Nowak, G. Papp and I. Zahed, Nucl. Phys. {\bf B501}, 603 (1997).

\bibitem{nuclear}J.J.M. Verbaarschot, H.A. Weidenm\"uller and M.R. Zirnbauer,
{\sl Grassmann integration in stochastic quantum physics: The case
of compound-nucleus scattering \/}, Phys. Rep. {\bf 129}, 367
(1985).\newline
Y.V. Fyodorov and H.-J. Sommers, {\sl Statistics of resonance poles,
phase shifts and time delays in quantum chaotic scattering:
Random matrix approach for systems with broken time-reversal invariance \/},
J. Math. Phys. {\bf 38}, 1918 (1997); {\sl Random Matrices
Close to Hermitean or Unitary: Overview of Methods and Results.
\/}, J. Phys. A {\bf 36}, 3303 (2003). (Special Issue on ``Random
Matrix theory".)\newline
Y. V. Fyodorov, B. A. Khoruzhenko and H.-J. Sommers, Phys. ~Rev. ~Lett.
{\bf 79 } (1997) 557; Phys. Lett A {\bf 226} (1997) 46; 
 Ann.~Inst.~H.~Poincare {\bf 68 } (1998) 449.
\newline
Y. V. Fyodorov, {\sl Almost-Hermitean Random Matrices: Applications to the 
Theory of Quantum Chaotic Scattering and Beyond}, published in 
{\em Supersymmetry and Trace Formulas: Chaos and Disorder}, I. Lerner et al. 
(Editors), Kluwer Academic/Plenum Publishers, NY 1999, p. 293.
\newline
E. Gudowska-Nowak, G. Papp and J. Brickmann, cond-mat/9701187.


\bibitem{nuclear1}F. Haake, F. Izrailev, N. Lehmann, D. Saher and H. J.
Sommers, Zeit. Phys. B {\bf 88} (1992) 359. ~H. J. Sommers, A. Crisanti,
H. Sompolinsky and Y. Stein, Phys. ~Rev.~Lett. {\bf 60} (1988) 1895.


\bibitem{hn} N. Hatano and D. R. Nelson, Phys. ~Rev. ~Lett. 
{\bf 77 } (1997) 570; Phys. ~Rev. {\bf B 56} (1997) 8651; Phys. ~Rev. 
{\bf B 58} (1998) 8384.
\newline
N. Hatano, {\sl Localization in Non-Hermitean Quantum Mechanics and Flux-Line 
Pinning in Superconductors}, Physica {\bf A 254} (1998) 317 
(cond-mat/9801283).
\newline
A. Zee, {\sl A Non-Hermitean Particle in a Disordered World}, Physica 
{\bf A 254} (1998) 300 (cond-mat/9711114).

\bibitem{efetov} K. B. Efetov, Phys. ~Rev. ~Lett. {\bf 79 } (1997) 491; 
Phys. ~Rev. {\bf B 56} (1997) 9630.

\bibitem{resonances} T. Kottos, 
{\sl Statistics of Resonances and Delay Times in Random Media:
Beyond Random Matrix Theory\/}, J. Phys. A {\bf 38} (2005) 10761 
(Special Issue on "Aspects of Quantum Chaotic Scattering", cond-mat/0508173). 
\newline
M. Rusek, J. Mostowski and A. Orlowski, Phys. Rev. A
{\bf 61}, 022704 (2000).

\bibitem{chalker} J.T. Chalker and Z.J. Wang, Phys. Rev. Lett. 
{\bf 79}, 1797 (1997); Phys. Rev. {\bf E 61}, 196 (2000).


\bibitem{encyclopedia} M.A. Stephanov, J.J.M. Verbaarschot and T. Wettig, 
{\sl Random Matrices}, published in the Wiley Encyclopedia of Electrical 
and Electronics Engineering, Supplement 1 (2001), hep-ph/0509286. 

\bibitem{BIPZ} E. Br\'ezin, C. Itzykson, G. Parisi and J.~-B.
Zuber, Comm.~Math.~Phys. {\bf 59} (1978) 35.

\bibitem{ginibre} J. Ginibre, Jour. ~Math. ~Phys. {\bf 6} (1965) 440.


\bibitem{diagrams} D. Bessis, C. Itzykson and J.-B. Zuber, Adv.~Appl.~Math.
~{\bf 1} (1980) 109.\newline
For more recent references, see e.g., E.~Br\'ezin and A. ~Zee, Phys. ~Rev. 
{\bf E 49} (1994) 2588.
\newline
A. ~Zee, {\sl Quantum Field Theory in a Nutshell}, Princeton
Univ. Press, Princeton, 2003. (Chapter VII.4.) 

\bibitem{rg} E.~Br\'ezin and A. Zee, Comp. ~Rend. ~Acad. 
~Sci., (Paris)  {\bf 317} (1993) 735. ~E. ~Br\'ezin and J. ~Zinn-Justin, 
Phys. Lett B {\bf 288} (1992) 54.\newline
S. ~Higuchi, C.~Itoh, S.~Nishigaki and N.~Sakai,  ~Phys.~Lett. {\bf B 318} 
(1993) 63; ~Nucl. ~Phys. {\bf B 434} (1995) 283, Err.-ibid. {\bf B 441} 
(1995) 405.\newline
J.~D'Anna and A. ~Zee, Phys. ~Rev. {\bf E 53} (1996) 1399.~
J.~Feinberg and A. ~Zee, Jour. ~Stat. ~Phys. {\bf 87} (1997) 473. 

\bibitem{chiral} There are many papers on chiral matrices. 
A partial list:\\
E.~Br\'ezin, S.~Hikami and A.~Zee,  Phys. ~Rev. {\bf E 51}
(1995) 5442;~Nucl. ~Phys.{\bf B 464} (1996) 411\,.\newline
J.~J.~M.~Verbaarschot, Nucl.~Phys. {\bf B426} (1994) 559; ~~J.~J.~M.~
Verbaarschot and I.~Zahed, Phys.~Rev.~Lett. {\bf 70} (1993) 3852.\newline
G.M. Cicuta and E. Montaldi, Phys.~Rev.~ {\bf D 29} 
(1984) 1267;~~A. Barbieri, G.M. Cicuta and E. Montaldi, Nuov.~Cim.~ 
{\bf 84 A} (1984) 173;~~C.M. Canali, G.M. Cicuta, L. Molinari and E. 
Montaldi, Nucl. ~Phys. {\bf B 265}, (1986) 485;~~G.M. Cicuta, L. 
Molinari, E. Montaldi and F. Riva,  J. ~Math. ~Phys. {\bf 28} (1987) 
1716.\newline  
K.~Slevin and T.~Nagao, Phys.~Rev.~Lett. {\bf 70} (1993) 635, Phys.~Rev.~ 
{\bf B 50 } (1994) 2380;~~T.~Nagao and K.~Slevin, J. ~Math. ~Phys. {\bf 34} 
(1993) 2075, 2317.\newline
T. ~Nagao and P. ~J. ~Forrester, Nucl.~Phys. {\bf B 435} (FS) (1995)
401.\newline
A.~V.~Andreev, B.~D.~Simons and N.~Taniguchi, Nucl.~Phys. {\bf B 432} (1994) 
485.\newline
S. Hikami and A. Zee, Nucl. ~Phys. {\bf B 446}, (1995) 337;~~S.~Hikami, 
M.~Shirai and F.~Wegner, Nucl.~Phys. {\bf B 408} (1993) 415.\newline
C.B.~Hanna, D.P. ~Arovas, K. ~Mullen and S.M.~Girvin, cond-mat 9412102.

\bibitem{chiral1} A.~ Anderson, R.~C. Myers and V.~ Periwal, Phys. Lett.
{\bf B 254} (1991) 89,~~ Nucl. ~Phys. {\bf B 360}, (1991)  463.\newline
R.~C. Myers and V.~ Periwal, Nucl. ~Phys. {\bf B 390}, (1991) 716.\newline
J.~Ambj{\o}rn, J.~Jurkiewicz, and Yu.~M.~Makeenko, Phys.~Lett. {\bf B251} 
(1990) 517;~~S. ~Nishigaki  Phys.~ Lett. {\bf B 387} (1996) 139.

\bibitem{ambjorn} J.~Ambj{\o}rn, {\sl Quantization of Geometry}, in Les 
Houches 1990, edited by J.~Dalibard et al. See section 4.3 and references 
therein.

\bibitem{thooft} G. ~'t~Hooft, Nucl.~Phys. {\bf B 72}, (1974) 461. 

\bibitem{coleman} S. Coleman, {\sl Aspects of Symmetry} 
(Selected Erice Lectures), Cambridge Univ. Press, Cambridge 1995, Chapter 8.3. 

\bibitem{BN} E.~Br\'ezin and H.~Neuberger,  Nucl.~Phys. {\bf B 350}, (1991) 
513. 

\bibitem{replica} As an introductory reference on the replica trick 
see e.g. A. Zee's book cited in \cite{diagrams}, Chapter VI.7.

\bibitem{bluez}
A.~Zee, Nucl.~Phys. {\bf B474} (1996) 726.


\end{thebibliography}
\end{document}